\documentclass[prd,superscriptaddress,showpacs,nofootinbib,amsmath,amssymb]{revtex4}

\usepackage{bm}
\usepackage{amsfonts}
\usepackage{latexsym}
\usepackage[latin1]{inputenc}
\usepackage{graphicx}
\usepackage{amsmath}
\usepackage{rotating}
\usepackage{epsfig}

\newcommand{\ep}{\epsilon}
\newcommand{\om}{\omega}
\newcommand{\Om}{\Omega}

\newcommand{\la}{\lambda}
\newcommand{\veps}{\varepsilon}

\newcommand{\II}{\mbox{\rm i}}
\newcommand{\U}{\textrm{U}}
\newcommand{\Sv}{\textrm{S}}
\newcommand{\V}{\textrm{V}}
\def \nn  {\nonumber}

\def \eps{\epsilon}
\def \veps{\varepsilon}
\def\jnl@style{\it}
\def\aaref@jnl#1{{\jnl@style#1}}

\def\aaref@jnl#1{{\jnl@style#1}}

\def\aj{\aaref@jnl{AJ}}                   
\def\apj{\aaref@jnl{ApJ}}                 
\def\apjl{\aaref@jnl{ApJ}}                
\def\apjs{\aaref@jnl{ApJS}}               
\def\apss{\aaref@jnl{Ap\&SS}}             
\def\aap{\aaref@jnl{A\&A}}                
\def\aapr{\aaref@jnl{A\&A~Rev.}}          
\def\aaps{\aaref@jnl{A\&AS}}              
\def\mnras{\aaref@jnl{MNRAS}}             
\def\prd{\aaref@jnl{Phys.~Rev.~D}}        
\def\prl{\aaref@jnl{Phys.~Rev.~Lett.}}    
\def\qjras{\aaref@jnl{QJRAS}}             
\def\skytel{\aaref@jnl{S\&T}}             
\def\ssr{\aaref@jnl{Space~Sci.~Rev.}}     
\def\zap{\aaref@jnl{ZAp}}                 
\def\nat{\aaref@jnl{Nature}}              
\def\aplett{\aaref@jnl{Astrophys.~Lett.}} 
\def\apspr{\aaref@jnl{Astrophys.~Space~Phys.~Res.}} 
\def\physrep{\aaref@jnl{Phys.~Rep.}}      
\def\physscr{\aaref@jnl{Phys.~Scr}}       

\begin{document}

\title[Differential Rotation]{Non-axisymmetric oscillations of
differentially rotating relativistic stars}


\author{Andrea Passamonti} \email{passamonti@astro.auth.gr}
\affiliation{Department of Physics, Aristotle University of Thessaloniki, 54124, Greece} 

\author{Adamantios Stavridis\footnote{ Present address : Groupe de
Gravitation et Cosmologie (GR$\varepsilon$CO), Institut d'
Astrophysique de Paris (CNRS), 98 Boulevard Arago, 75014, Paris,
France} } 
\email{astavrid@astro.auth.gr} \affiliation{Department of
Physics, Aristotle University of Thessaloniki, 54124, Greece}

\author{Kostas D. Kokkotas} 
\affiliation{Department of Physics, Aristotle University of
Thessaloniki, 54124, Greece} 
\affiliation{Theoretical Astrophysics,
University of T\"ubingen, Auf der Morgenstelle 10, T\"ubingen 72076,
Germany}
\date{\today}


\begin{abstract}
Non-axisymmetric oscillations of differentially rotating stars are
 studied using both slow rotation and Cowling approximation.  The
 equilibrium stellar models are relativistic polytropes where
 differential rotation is described by the relativistic j-constant
 rotation law.  The oscillation spectrum is studied versus three main
 parameters: the stellar compactness $M/R$, the degree of differential rotation
 $A$ and  the number of maximun couplings $\ell_{\rm max}$. It is
 shown that the rotational splitting of the non-axisymmetric modes is strongly
 enhached by increasing the compactness of the star and the degree of differential
 rotation.  Finally, we investigate the relation between the
 fundamental quadrupole mode and the corotation band of differentially
 rotating stars. 

\end{abstract}


\maketitle

\section{Introduction}
Differential rotation is believed to play an important role in nascent
neutron stars \cite{Dimmelmeier:2002bm} as well as in binary neutron
stars mergers \cite{Rasio:1994bd,Shibata:1999wm}. Until viscosity,
turbulent motion and/or magnetic fields force the star to uniform
rotation, dynamical or secular instabilities can be developed due to 
differential rotation.

The study of the differential rotation phase of a neutron star life
re-gained attention lately. Recent numerical simulations in Newtonian
hydrodynamics
\cite{2001ApJ...550L.193C,2002MNRAS.334L..27S,2005ApJ...625L.119O,2006ApJ...651.1068O}
show that in stars with high differentially rotating, non-axisymmetric
dynamical instabilities can develop even for low values of
$\beta=T/|W| \simeq 0.01-0.08$, where $T$ is the rotational kinetic
energy and $W$ the gravitational binding energy. The so-called ``low
$T/|W|$'' instability can drive either one-armed spiral or bar-mode
instabilities.  The gravitational wave signal emitted via these
instabilities might be detectable with the advanced generation of
ground based detectors.  If this instability develops in supermassive
stars \cite{2001ApJ...548..439N}, it may produce gravitational waves
detectable even by the interferometric space detector LISA.  As
suggested in~\cite{2005ApJ...618L..37W}, the low $T/|W|$ instability
might be due to the shear instability that the corotating $f$ mode
develops when it enters the corotation band.  A Newtonian study of
this instability~\cite{2006MNRAS.368.1429S}, which is based on the
analysis of the canonical angular momentum, confirms the presence of
corotation modes.

A key feature of the oscillation spectrum of a uniformly or
differentially rotating star is the splitting of the eigenmodes, like
the Zeeman effect in atomic physics. In Newtonian theory, the
splitting of the non-axisymmetric oscillations of differentially
rotating stars have been studied with perturbative techniques in
\cite{1977ApJ...217..151H}.  A general relativistic description of the
perturbations of differentially rotating neutron stars may alter, at
least quantitatively, the Newtonian results.  In general relativity,
the perturbations of uniformly rotating stars are mainly studied in
the so called slow rotation approximation. This approximation is
practically valid for the study of all known pulsars even for those
with rotational periods of 1.5-2ms. The slow rotation approximation
fails to treat perturbations of neutron stars with periods below 1.5ms
or $\Omega/\Omega_{\rm Kepler}>0.25-0.3$.  Actually, there are no
known pulsars with such small periods but it is not at all impossible
that newly born neutron stars may have rotational periods below 1ms.

The lack of perturbative studies for differentially rotating relativistic stars, the absence of any results for non-axisymmetric perturbations and the issue of low $T/|W|$ instability
are the main motivations of this work. There only a few studies of fast rotating neutron stars using a perturbative approach
\cite{Yoshida:2004gk,Boutloukos:2006cx,2007CQGra..24.4147S,2007arXiv0709.2925F}. While there is significant progress in the study of neutron star
oscillations using evolutions of the non-linear equations
\cite{Font:1999wh,Stergioulas:2003ep,Dimmelmeier:2005zk}. Still the
non-axisymmetric perturbations of differentially rotating relativistic
stars have not been treated by any method yet, this paper is the first
attempt to address the problem.
 Actually, in an earlier paper we derived,
in the perturbative framework of general relativity, the equations
describing the oscillations of a slowly and differentially rotating
neutron star in the Cowling approximation \cite{2007PhRvD..75f4019S}
(from now on Paper I).  Here we study the effect of differential
rotation on the oscillation spectrum.
Moreover, by using the perturbative approach
developed here we examine the relation between the low $T/|W|$
instability with the existence of corotating modes.  The main results
of our study can be summarized in the following sentence: \emph{the
rotational splitting of the non-axisymmetric modes is enhanced by stellar compactness
and the degree of differential rotation}.

The structure of this paper is as follows. In section
\ref{Sec:Pert.Frame} we briefly describe the perturbative framework
(more details are given in Paper I).  In Section \ref{sec:pert-eqs} we
address the boundary value problem and introduce the numerical
techniques for solving it. In Section \ref{Sec:Num.Results} we present
and discuss our numerical results derived for different stellar
models. Section~\ref{sec:concl} is dedicated to the conclusions and to
the possible extensions of this work.  Finally, in the
Appendix~\ref{app:matrix}, we describe the structure of the eigenvalue
equations, while the definition of the linear angular operators is
given in Appendix~\ref{Integrals}.

As is common, throughout the paper we use geometrical units
i.e. $c=G=1$.  Prime (~$'$~) denotes derivatives with respect to the
radial coordinate $r$ and overdot (~$\dot{}$~) denotes derivatives
with respect to the time coordinate $t$.

\section{The Perturbative Framework}
\label{Sec:Pert.Frame}
Equilibrium configurations of differentially rotating relativistic
stars have been already studied in the early '70s by Hartle
\cite{Hartle:1970ha} and Will \cite{1974ApJ...190..403W}.  In the
approach that we follow here, the background spacetime of a slowly and
differentially rotating star assumed to be axially symmetric and can
be described, at first order with respect to the angular velocity of
the star, in Schwarzschild coordinates, by the following line element:
\begin{align}
ds^2 = - e^{2\nu} dt^2 + e^{2\la} dr^2 & - 2 \, \omega \, r^2 \sin ^2
    \theta \, dt \, d \phi + r^2 \left( d \theta ^2 + \sin^2 \theta d
    \phi ^2 \right) \, . \label{ds-equil}
\end{align}
The scalar functions $\nu$, $\la$ depend only on the radial
coordinate $r$ and are determined by solving the
Tolman-Oppenheimer-Volkoff~(TOV) equations for a given equation of
state~(EoS).  The metric function $\omega=\omega\left( r,\theta
\right)$ describes the dragging of inertial frames due to stellar
rotation and obeys the following ODE~\cite{Hartle:1967ha}:
\begin{widetext}
\begin{eqnarray}
\omega^{''} - \left[ 4 \pi \left( \ep + p \right) r e^{2\la} -
  \frac{4}{r} \right] \omega^{'} - \left[ 16 \pi \left( \ep + p
  \right) + \frac{\ell\left(\ell+1\right) - 2 }{r^2} \right] e^{2\la} \omega
= - 16 \pi \left( \ep + p \right) e^{2\la} \Omega \, ,
\label{drag-eq}
\end{eqnarray}
\end{widetext}
where $\ep$ and $p$ are the total energy density and the pressure
respectively, $\ell$ the harmonic index and
$\Omega=\Omega\left(r,\theta \right)$ is the angular velocity of the
star as measured by an observer at infinity.

A differentially rotating stellar model can be constructed in the
framework of the slow rotation approximation in two steps.  First, one
constructs the non-rotating stellar model by specifying the central
energy density and the EoS and then solving the TOV equations.
Afterwards, a law describing the differential rotation is specified,
that is one assumes a specific functional form for $\Omega(r,\theta)$
and then the metric function $\omega(r,\theta)$ is determined by
solving ODE~(\ref{drag-eq}).
Here, we use the perturbative version of the relativistic
j-constant rotation law~\cite{Passamonti:2005cz, 2007PhRvD..75f4019S}:
\begin{equation}
\Omega(r,\theta) = \frac{A^2\Omega^{}_{c} + e^{-2\nu} \omega(r,\theta) r^2 \sin^2\theta\,}
{A^2 + e^{-2\nu} r^2 \sin^2\theta} \, , \label{j-cons}
\end{equation}
where $\Omega_c$ denotes the angular velocity on the rotation axis,
while the parameter $A$ specifies the degree of differential rotation
of the star. For $A\to \infty$, the j-constant rotation
law~(\ref{j-cons}) leads to  uniformly rotating configurations, i.e.
$\Omega \rightarrow \Omega_c$. More details about the procedure of constructing
equilibrium configurations are given in Paper I, where we have
adopted a harmonic expansion of the variables $\omega$ and $\Omega$ up
to  $\ell=3$.

The perturbation equations describing non-barotropic, non-axisymmetric
oscillations of slowly and differentially rotating relativistic stars
have already been derived in Paper I. We have actually used the so
called Cowling approximation, i.e. the equations are derived by
perturbing only the fluid variables in the energy momentum equations
$\delta \left( T_{\mu\nu}^{\;\;\;\; ; \mu} \right) = 0$.  The
oscillations of the fluid are then described by five functions, namely
the enthalpy $H^{\ell m}$, the three perturbed velocity components
$u_{1}^{\ell m}$, $u_{2}^{\ell m}$ (polar), $u_{3}^{\ell m}$ (axial)
and the radial component of Lagrangian displacement vector $\xi^{\ell
m}$.  The perturbation equations read:
\begin{widetext}
\begin{eqnarray}
\label{Eq_H}
 \dot{H}^{\ell m} + \II m \left( \Om _1 + 6 \Om _3 \right) H^{\ell m} &
                  = & \left\{ \left[ \left( \frac{2}{r} - \la ' + 2
                  \nu' \right) c_s^2 - \nu ' \right] u_{1}^{\ell m}
                  + c_s^2 (u_{1}^{\ell m}) ' \right\} e^{2\left( \nu -
                  \la \right) } - c_s^2 \Lambda \frac{e^{2\nu}}{r^2}
                  u_{2}^{\ell m} \nn \\ \nn \\ & + & \II m \left\{ 2
                  c_s^2 \left( \varpi_1 + 6 \varpi_3 \right) -
                  \frac{15}{2} \left( 2 c_s^2 \varpi_3 - \Om _3
                  \right) \mathcal{L} _{1}^{\pm 2} \right\} H^{\ell m}
                  \, ,
\end{eqnarray}
\begin{eqnarray}
\label{Eq_u1}
\dot{u}_{1}^{\ell m} +  \II m \left( \Om _1 + 6 \Om _3 \right) u_{1}^{\ell m} & = &
                         (H^{\ell m})'  + \nu' c_s^{-2} \left[ - \xi^{\ell m} +
               \left( 1 - \frac{\Gamma_1}{\Gamma} \right) H^{\ell m} \right]
                     + \frac{15}{2} \II m \Om _3  \, \mathcal{L}  _{1}^{\pm 2}  u_{1}^{\ell m} \nn \\ \nn \\
                 & + & \II m \left\{
                      \left[ 2 \left( \frac{1}{r} - \nu' \right) \varpi_1 - \omega_1' \right]
                   +   \left[ 2 \left( \frac{1}{r} - \nu' \right)\varpi_3 - \omega_3' \right]
                     \left( 6 - \frac{15}{2} \, \mathcal{L}  _{1}^{\pm 2}\right) \right\} u_{2}^{\ell m}
                         \nn \\  \nn \\
                 & + &  \left\{ \left[ 2 \left( \frac{1}{r} - \nu' \right) ( \varpi_1 + 6 \varpi_3 )
                         - \omega_1' - 6 \omega_3' \right]  \mathcal{L}  _{1}^{\pm 1}  \right. \nn \\ \nn \\
                 & - & \left. \frac{15}{2} \left[ 2 \left( \frac{1}{r} - \nu' \right) \varpi_3 - \omega_3' \right]
                      \mathcal{L}  _{1}^{\pm 3}   \right\} u_{3}^{\ell m} \, ,
\end{eqnarray}
\begin{eqnarray}
\label{Eq_u2}
\dot{u}_{2}^{\ell m}  + \II m \left( \Om _1 + 6 \Om _3  \right) u_{2}^{\ell m}  & = &
                             H^{\ell m}  \nn \\ \nn \\
                           & + & \frac{\II m}{\Lambda}  r e^{-2\lambda} \left\{   \left( 2 - 2 r \nu ' \right)
                               \left( \varpi_1 + 6 \varpi_3 \right)
                             + r ( \varpi_1' + 6 \varpi_3' )  \right. \nn \\ \nn \\
                           & - & \left. \frac{15}{2} \left[ \left( 2 - 2 r \nu ' \right) \varpi_3
                               + r \varpi_3' \right]  \mathcal{L}  _{1}^{\pm 2}
                                 \right\}  u_{1}^{\ell m}  \nn \\ \nn \\
                           & + & \frac{\II m}{\Lambda}
                                  \left\{ 2 ( \varpi_1 + 6 \varpi_3 )
                               - 30 \varpi_3 \mathcal{L} _{3}^{\pm 2} - 15 ( \Om_3 - 2\om_3 ) \mathcal{L} _{2}^{\pm 2}
                                 + \frac{15}{2}  \Om_3 \mathcal{L} _{4}^{\pm 2} ) \right\} u_{2}^{\ell m} \nn \\ \nn \\
                           & + & \frac{1}{\Lambda} \left\{ 2 ( \varpi_1 + 6 \varpi_3 ) \mathcal{L} _{3}^{\pm 1}
                                  - 15 m^2 ( 2 \varpi_3 + \Om_3  ) \mathcal{L} _{4}^{\pm 1}
                                  - 15 ( \Om_3 - 2 \om_3 ) \mathcal{L}
                                  _{2}^{\pm 3} \right\}  u_{3}^{\ell m}
                                \, ,
\end{eqnarray}
\begin{eqnarray}
\label{Eq_u3}
\dot{u}_{3}^{\ell m}  +  \II m \left( \Om _1 + 6 \Om _3  \right) u_{3}^{\ell m}
                 & = & \frac{\II m}{\Lambda} \left\{ 2 ( \varpi_1 + 6 \varpi_3 )
                                 - 30 \varpi_3 \mathcal{L}_{2}^{\pm 2}
                                 - 15 ( \Om_3 - 2\om_3 )  \mathcal{L}_{3}^{\pm 2}
                                 + \frac{15}{2} \Om_3 \mathcal{L}_{4}^{\pm 2} \right\} u_{3}^{\ell m} \nn \\ \nn \\
                 & - & \frac{1}{\Lambda} \left\{ 2 (\varpi_1 + 6 \varpi_3 ) \mathcal{L}_{3}^{\pm 1}
                              - 30 m^2 \varpi_3  \mathcal{L}_{4}^{\pm 1} - 30 \varpi_3 \mathcal{L}_{2}^{\pm 3}
                                                 \right\} u_{2}^{\ell m} \nn \\ \nn \\
                 & - & \frac{r}{\Lambda} \,  e^{-2\lambda}
                               \left\{ \left(  \left( 2 - 2 r \nu' \right)  ( \varpi_1 + 6\varpi_3 )
                            + r ( \varpi_1' + 6 \varpi_3' ) \right] \mathcal{L}_{2}^{\pm 1} \right. \nn \\ \nn \\
                 & - &  \left. \frac{15}{2} \left[
                        \left( 2 - 2 r \nu' \right) \varpi_3
                        + r \varpi_3' \right] \mathcal{L}_{3}^{\pm 3}
                    \right\}  u_{1}^{\ell m} \, , \\
\label{Eq_xi}
\dot{\xi}^{\ell m} + \II m \left( \Om_{1} + 6 \Om_{3} \right ) \xi^{\ell m}
& = & e^{2\nu-2\lambda} \left( \frac{\Gamma_1}{\Gamma} - 1\right) \nu' u_{1}^{\ell m}
 +  \frac{15}{2} \II m \Om_{3} \mathcal{L}_1^{\pm 2} \xi^{\ell m} \, ,
\end{eqnarray}
\end{widetext}
where $\varpi = \Omega - \omega$. The explicit form of the
linear angular operators $\mathcal{L}_{i}^{\pm j}$ are given in
Appendix~\ref{Integrals}. 

As in the case of uniformly rotating
stars~\cite{2003MNRAS.339.1170R,2005IJMPD..14..543S},
Eqs.~(\ref{Eq_H}-\ref{Eq_xi}) form an infinitely coupled system of
equations.  In particular, differential rotation introduces extra
couplings with respect to the uniformly rotating case.
In the limit of uniform stellar rotation, $\Omega_3$ and $\omega_3$
vanish and Eqs.~(\ref{Eq_H}-\ref{Eq_xi}) reduce to the perturbative
equations presented in~\cite{2003MNRAS.339.1170R}.  In addition, like
in the uniformly rotating case, these equations can be devided into
two independent subsystems, the so called axial-led and polar-led
\cite{Lockitch:2000aa}.  The polar-led system is the one that includes
polar perturbations with $\ell=|m| + 2 k$, and axial perturbations
with $\ell=|m| + 2 k + 1$, where $k$ is an integer.  Instead, the
axial-led system is the one that includes axial perturbations with
$\ell=|m| + 2 k$, and polar perturbations with $\ell=|m| + 2 k + 1$.
The overall parity of each system is preserved and is polar for the
first and axial for the second. In the rest of the paper we will focus
mainly on the polar-led perturbations and leave axial-led for another
study.  Finally, for barotropic oscillations, where the background
adiabatic index $\Gamma$ is equal to the adiabatic index of the
perturbations $\Gamma_1$, the last equation is obsolete and the system
reduces to four evolution equations.

The study of the oscillations described by
Eqs.~(\ref{Eq_H}-\ref{Eq_xi}) can be done either in the time domain as
an initial value problem or in the frequency domain as an eigenvalue
problem~(Sec.~\ref{sec:pert-eqs}). Our study was based on the last
method although we have also tried numerical evolutions of the system.
In this case, we implemented a numerical code based on the two step
Lax Wendroff scheme~\cite{Thomas:1995}. For some of the stellar models, the simulations
show some numerical instabilities after an evolution time of $\simeq
20-30~{\rm ms}$.  Before the developement of the instability, we were
able to determine with a Fast Fourier Transformation (FFT) the mode
frequencies which agree better than $\simeq 3-5 \%$ to the ones
derived by an eigenvalue problem.

\section{Perturbation equations in frequency domain} \label{sec:pert-eqs}

The perturbation equations~(\ref{Eq_H}-\ref{Eq_xi}) can be studied in
the frequency domain by assuming a harmonic time dependence for the
perturbation functions $e^{- \II \sigma t}$, with $\sigma$ being the
oscillation frequency. By replacing the time derivatives in
Eqs.~(\ref{Eq_H}-\ref{Eq_xi}) by $-\II \sigma$ (e.g. ${\dot H}\to -
\II \sigma H$) and transforming $H^{\ell m} \to \II H^{\ell m}$,
$u_3^{\ell m} \to \II u_3^{\ell m}$ and $\xi^{\ell m} \to \II
\xi^{\ell m}$ one obtains a purely real eigenvalue problem, which is
formed by two ODEs for $H^{\ell m}$ and $u_1^{\ell m}$ and three
algebraic equations for $u_2^{\ell m}$, $u_3^{\ell m}$ and $\xi^{\ell
m}$.

In order to prescribe the eigenvalue problem in a more compact form we
first define the following three vectors:
\begin{eqnarray}
u^{\ell m} \equiv \left( H^{\ell m}, u_1^{\ell m} \right)^{\rm T} \, ,
\qquad \qquad s^{\ell m} \equiv \left( u_2^{\ell m}, u_3^{\ell m}
\right)^{\rm T} \, , \qquad \qquad v^{\ell m} \equiv \left( \xi^{\ell m}, 0
\right)^{\rm T} \, , \label{def:usv}
\end{eqnarray}
and then three infinite dimensional vectors:
\begin{eqnarray}
\U \equiv  \left( \dots, u^{\ell-1 m} ,u^{\ell m}, u^{\ell+1 m}, \dots \right)^{\rm T} \, , \quad
\Sv \equiv \left( \dots, s^{\ell-1 m}, s^{\ell m}, s^{\ell+1 m}, \dots \right)^{\rm T} \, , \quad
\V \equiv \left( \dots, v^{\ell-1 m},, v^{\ell m}, v^{\ell +1 m}, ..~
\right)^{\rm T} \, .   \label{def:USVvec}
\end{eqnarray}
By using these definitions, the perturbative equations can be written in an operatorial form
as follows:
\begin{eqnarray}
\label{EqdUdr}
   \frac{d\mathbf{\U}}{dr}   & = & \mathcal{A}_{\sigma}^{\rm tot} \cdot \U  +
                                   \mathcal{C} \cdot \Sv +
                       \mathcal{D} \cdot \V \, ,  \label{eq:u}  \\  \nn \\
\label{EqS}
   \mathcal{S}_{\sigma} \cdot \Sv & = &  \mathcal{M} \cdot \U \, ,
   \label{eq:s}  \\  \nn \\
\label{EqV}
   \mathcal{Q}^{\rm tot}_{\sigma} \cdot \V  & = &   \mathcal{N} \cdot \U   \, ,  \label{eq:xi}
\end{eqnarray}
where $\mathcal{A}_{\sigma}^{\rm tot}$, $\mathcal{C}$, $\mathcal{D}$,
$\mathcal{S}_{\sigma}$, $\mathcal{M}$, $\mathcal{N}$,
$\mathcal{Q}^{\rm tot}_{\sigma}$, are infinite dimensional linear
operators which are defined in Appendix~\ref{app:matrix}.  In
$\mathcal{A}^{\rm tot}_{\sigma}$, $\mathcal{S}_{\sigma}$,
$\mathcal{Q}^{\rm tot}_{\sigma}$ the subscript denotes the dependence
of these matrices on the oscillation frequency $\sigma$.  These
operators, as already mentioned above, couple perturbations with
different harmonic indices $\ell$, leading to an infinite system of
coupled equations.  For a given azimuthal index $m$, the number of
couplings in the system of equations~(\ref{eq:u}-\ref{eq:xi}) can be
controlled by the parameter $\ell_{\rm max}$, which is the harmonic
index $\ell$ where the tensor harmonic expansion of perturbations is
truncated.  In this way, the number of couplings is $n_c = \ell_{\rm
max} - |m| +1$ as $\ell$ runs $|m|$ to $\ell_{\rm max}$ i.e. $|m| \le
\ell \le \ell_{\rm max}$.  The truncation of the couplings up to a
certain $\ell_{\rm max}$ has been done for practical reasons since it
is impossible to deal with infinity many couplings which do not
contribute significantly in the final result.  This approximation has
been tested by studying the variation of the eigenfrequencies with
respect to a varying value of $\ell_{\rm max}$. In general, we found
that even by keeping only a small number of couplings the
eigenfrequencies were converging~(Sec.~\ref{sec:spectrum}).

One can form a boundary value problem by using
Eqs.~(\ref{eq:u}-\ref{eq:xi}) together with an appropriate set of
boundary conditions at the center and the surface of the star. In the parameter
domain where the inverse of the operators $\mathcal{S}_{\sigma}$ and
$\mathcal{Q}^{\rm tot}_{\sigma}$ exist, both Eqs.~(\ref{EqS}) and (\ref{eq:xi})
can be solved for $\Sv$ and $\V$ respectively:
\begin{equation}
\Sv  =   \mathcal{S}_{\sigma}^{-1} \cdot \mathcal{M} \cdot \U \, ,   \qquad \qquad
\V   =   \left( \mathcal{Q}^{\rm tot}_{\sigma} \right)^{-1} \cdot
\mathcal{N} \cdot \U   \, , 
 \label{eq:inveq}
\end{equation}
and inserted into Eq.~(\ref{eq:u}). Then the final matrix
equation, together with the appropriate boundary conditions, 
can  be used for the calculation of the eigenfrequency $\sigma$. 

In the domain of frequencies where the two operators
$\mathcal{S}_{\sigma}$ and $\mathcal{Q}^{\rm tot}_{\sigma}$ are
singular, Eqs.~(\ref{EqS}) and (\ref{EqV}) cannot be solved for $\Sv$
and $\V$. 
This issue will be discussed in more detail in Sec.~\ref{sec:contspec}.

\subsection{Boundary Conditions} \label{sec:bc}

The perturbation equations~(\ref{eq:u}-\ref{eq:xi}) can be solved as
an eigenvalue problem for the variable $\sigma$ by fixing the boundary
conditions at the center and the surface of the star. Regularity at
the center of the star suggests that the various perturbation
functions have the following behavior:
\begin{equation}
H^{\ell m} \sim r^{\ell} \, , \qquad u_1^{\ell m} \sim r^{\ell -1} \, ,
\qquad u_2^{\ell m} \sim r^{\ell } \, , \qquad u_3^{\ell m} \sim r^{\ell +1} \, .
\end{equation}
For any combination of $\ell$ and $m$ there is only one independent solution at the
center, as $H^{\ell m}$ and $u_1^{\ell m}$ are related via the relation:
\begin{equation}
 \left.  \left[ \ell \, H^{\ell m}  - \left( - \sigma + m \Omega_1
  \right)  u_1^{\ell m}  \right] \right| _{r=0} = 0\,
 . \label{eq:bcorigin}
\end{equation}
Moreover, since at the center of the star $H^{\ell m}$ and $H^{\ell'
m}$ are independent, there exist $\ell_{\rm max} - |m| + 1$
independent boundary conditions.

The second boundary condition comes from the vanishing of the
pressure's Lagrangian perturbation on the surface.  This condition is
satisfied when the perturbations obey the following $\ell_{\rm max} -
|m| + 1$ number of equations:
\begin{equation}
\left[ -\sigma + m \left( \Omega_1 + 6 \Omega_3 \right) - \frac{15}{2}
m ~\Omega_3 \mathcal{L}_{1}^{\pm 2} \right] e^{2\lambda-2\nu} H^{\ell
m} - \nu' u_1^{\ell m} = 0 \, , \qquad m \le \ell \le \ell_{\rm max} \,
. \label{eq:bc}
\end{equation}
Notice that in the previous equation, the operator acting on the enthalpy
perturbation $H^{\ell m}$ comes from the total derivative:
\begin{equation}
\frac{d}{dt} \equiv u^{\alpha} \nabla _{\alpha} = e^{-\nu} \left( \frac{\partial}{\partial t}
+ \Omega  \frac{\partial}{\partial \phi} \right)
\to \II e^{-\nu} \left[-\sigma + m \left( \Omega_1 + 6 \Omega_3 \right) - \frac{15}{2}
  m \sin^2\theta~\Omega_3 \right] \, . \label{def:totder}
\end{equation}
In fact, the last term of Eq.~(\ref{def:totder}) becomes the operator
$\mathcal{L}_{1}^{\pm 2}$ of Eq.~(\ref{eq:bc}) after the angular
integration of the perturbation equations.

\subsection{Numerical Method}
\label{}

The numerical technique that is used here has been already described
earlier in~\cite{2003MNRAS.339.1170R}.  By selecting a certain value
for $\ell_{\rm max}$ we allow $n_c = \ell_{\rm max} - |m| + 1$
couplings between various harmonics where $|m| \le \ell \le \ell_{\rm
max}$.  Then one must specify $n_c$ independent conditions at the
origin for $H^{\ell m}$ that can be denoted by a vector
\begin{equation}
\mathcal{H}_k \equiv \left(H^{\ell m},\dots,H^{\ell_{\rm max} m}  \right) 
\, , \mbox{where}\quad k=1,...,n_c,
\end{equation}
and here we assume that $\mathcal{H}_1 =\left( 1,0,\dots,0\right)$, 
$\mathcal{H}_2 = \left( 0,1,0,\dots,0\right), \dots ,\mathcal{H}_{n_c}
= \left( 0,\dots,0,1\right)$. Then the corresponding values for
$u_1^{\ell m}$ are estimated via Eq.~(\ref{eq:bcorigin}).  By
specifying a trial eigensolution $\sigma$, we integrate for each
$\mathcal{H}_k$ Eqs.~(\ref{eq:u}-\ref{eq:xi}) and we calculate the
value of the Lagrangian pressure perturbation $\Delta p_{k}^{\ell}$ on
the surface. This quantity in general does not vanish and after $n_c$
integrations, we construct an $n_c \times n_c$ ``pressure matrix''
$\mathbf{P}$.  An eigenmode $\sigma$ is a solution of
Eqs.~(\ref{eq:u}-\ref{eq:xi}) that simultaneously satisfy $\Delta
p^{\ell}=0$ for any $\ell$. Since the $n_c$ initial values for
$\mathcal{H}_k$ are independent, one can find appropriate coefficients
$a_k$ such that
\begin{equation}
\sum_{k=m}^{\ell_{\rm max}} a_k \Delta p_{k}^{\ell} = 0\, , 
\quad \mbox{for every} \quad \ell = m\dots \ell_{\rm max} \, .
\label{eq:sum_DP}
\end{equation}
 This is equivalent with the requirement that $\det \mathbf{P} = 0$.
After integrating Eqs.~(\ref{eq:u}-\ref{eq:xi}) for a wide range of
frequencies we isolate the zeros of $\det \mathbf{P}=0$ by a ``root
finding algorithm''. These zeros are the eigenmodes $\sigma$ of the
problem.

The eigenfunctions of the enthalpy and velocity perturbations can be
constructed as a linear combinations of the $n_c$ solutions that
correspond to the $n_c$ independent initial
conditions~$\mathcal{H}_k$. The coefficients $a_k$ of these linear
combinations are determined by solving the homogeneous system of
linear equations described by Eq~(\ref{eq:sum_DP}).

\subsection{Continuous Spectrum} \label{sec:contspec}

For the range of frequencies $\sigma$ that the inverse operators of
$\mathcal{S}_{\sigma}$ and $\mathcal{Q}_{\sigma}^{\rm tot}$ do not
exist, Eqs.~(\ref{eq:s}-\ref{eq:xi}) are singular and consequently the
eigenmodes cannot be determined using the frequency domain code. This
is the same singularity that generates the continuous spectrum as in
the r-mode studies~\cite{1998MNRAS.293...49K,
Beyer:1999te,2001MNRAS.328..678R, Ruoff:2002ta,2004CQGra..21.4661L},
which have been carried out in slow rotation approximation.  The
continuous spectrum (CS) practically can be determined by solving the
determinant of the $2n_c \times 2n_c$ matrices that represent the
operators $\mathcal{S}_{\sigma}$ and $\mathcal{Q}_{\sigma}^{\rm tot}$,
for the mode frequency $\sigma$.
Let us first consider the operator $\mathcal{S}_{\sigma}$ for the case
$\ell_{\rm max} = |m|$. The determinant is a second order polynomial in
$\sigma$ which has a double root of the form:
\begin{eqnarray}
\sigma & = & m \left( \Omega_1+6 \Omega_3 \right) - \frac{2 m}{\ell
 \left( \ell+1 \right)} \left( \omega_1 +6\,\omega_3 \right)
 -\frac{m}{\ell \left( \ell+1 \right)} \Psi
 \left(\ell,m,\Omega_3,\omega_3 \right) \, , \label{eq:Sconsp}
\end{eqnarray}
where the function $\Psi$ is defined as follows:
\begin{eqnarray}
\Psi & \equiv &  - 15 \left( 3 \Omega_3 - 4 \omega_3 \right) \left[ \ell  Q_{\ell+1,m}^2
- \left( \ell+1 \right) Q_{\ell,m}^{2} \right] \nn \\
& + & \frac{15}{2} \Omega_3  \left[
 \ell \left( \ell+1 \right) - \left( \ell+1 \right) \left( \ell-2 \right) Q_{\ell,m}^{2} - \ell \left( \ell+3 \right)
Q_{\ell+1,m}^{2} \right] \, , 
\end{eqnarray}
and the coefficient~$Q_{\ell m}$ is given in Eq.~(\ref{eq:Qlm}). The
interval of the CS is then $\sigma^S_R \le \sigma \le \sigma^S_c$,
where $\sigma^S_R$ and $\sigma^S_c$ are for the operator
$\mathcal{S}_{\sigma}$ the values of $\sigma$ given by
Eq.~(\ref{eq:Sconsp}) at stellar surface and centre respectively.
In general, for $\ell_{\rm max} \ge m$, there are $\ell_{\rm max} -
|m| + 1$ branches of the CS, which might as well overlap widening the
band of it.  Similarly to the uniformly rotating star
\cite{2003MNRAS.339.1170R}, for $\ell_{\rm max} \rightarrow \infty$
the CS will cover all the spectrum.

Similar behavior can be observed for the operator
$\mathcal{Q}_{\sigma}^{\rm tot}$, where one can calculate a singular
patch for $\ell_{\rm max} = m$ via the following expression:
\begin{equation}
\sigma = m \left( \Omega_1 + 6 \Omega_3 \right) -\frac{15}{2} m
 \Omega_3 \left[ 1- Q_{\ell,m}^2 - Q_{\ell+1,m}^2 \right] \,
 . \label{eq:Qconsp}
\end{equation}
Again in this case, the width of the CS is defined by the values
of $\sigma $ given by Eq.~(\ref{eq:Qconsp}) at the stellar center and
surface i.e. $\sigma^Q_R \le \sigma \le \sigma^Q_c$ and the number of
continuous spectrum patches when more coupling terms are included is
again $\ell_{\rm max} - |m| + 1$.

It is obvious from the above analysis, that the operators
$\mathcal{S}_{\sigma}$ and $\mathcal{Q}_{\sigma}^{\rm tot}$ generate
different bands of CS. For a barotropic EoS, Eq.~(\ref{eq:xi}) is
trivial and the continuous spectrum is only generated from operator
$\mathcal{S}_{\sigma}$.  Furthermore, when
expression~(\ref{eq:Qconsp}) is satisfied the output of the angular
integration of the total time derivative given by
Eq.~(\ref{def:totder}) vanishes and the CS of the operator
$\mathcal{Q}_{\sigma}^{\rm tot}$ lies in the corotation band of
oscillation modes.

It should be noted that the existence of a continuous spectrum for
uniformly rotating relativistic stars is highly debatable and it might
be a result of the slow rotation approximation \cite{2004CQGra..21.4661L}. 
On the other hand, differential rotation is known to be associated with the presence of
corotation points and existence of a continuous spectrum but this is
a technical problem that may not be solvable within the limits of the
slow rotation approximation. Thus we will not address the issues
related to the presence of CS in the rest of this paper.

\begin{table}[t]
\begin{center}
\begin{ruledtabular}
\begin{tabular}{*{6}{c}}
  Model  &  $\rho_c $  & $M$ &
         $R$  & $M/R$  & $\Omega_K$  \\
         &   $\left(\times 10^{-3}\textrm{km}^{-2} \right)$ &  $\left(M_{\odot}\right)$
         &   $\left(\textrm{km}\right)$   &  & $\left(\times 10^{-2}{\rm km}^{-1}\right)$ \\
\hline
  B0     &   0.662  & 1.400 &  14.151  & 0.146  &  2.700      \\
  C0     &   1.484  & 1.637 &  11.218  & 0.216  &  4.168      \\
\end{tabular}
\end{ruledtabular}
\caption{\label{tab:0Models} The parameters of the background non-rotating
 stellar models of the B and C sequence. Here $\rho_c$ is the central
 rest mass density, $M$ the mass, $R$ the radius and $\Om_K$ denotes
 the mass sheeding limit (Kepler frequency). }
\end{center}
\end{table}
\begin{table}[t!]
\begin{center}
\begin{ruledtabular}
\begin{tabular}{c | c c c c c }
  $a_k \left( \times 10 ^{-3}\right)$  &  $a_0 $ & $a_1$  & $a_2$ & $a_3$ & $a_4$  \\
\hline
  B     &  0.726  & -1.272 & 3.972 &-0.853 & 3.970  \\
  C     & -2.239  &  1.811 & 5.388 & 2.233 & 5.386  \\
\end{tabular}
\end{ruledtabular}
\caption{\label{tab:fitphi} The coefficients $a_k$ of the Pad\'e
approximation of Eq~(\ref{eq:Pade}) for both the B and C sequences of stellar modes.  }
\end{center}
\end{table}
\begin{figure}[ht]
\begin{center}
\includegraphics[width=85mm]{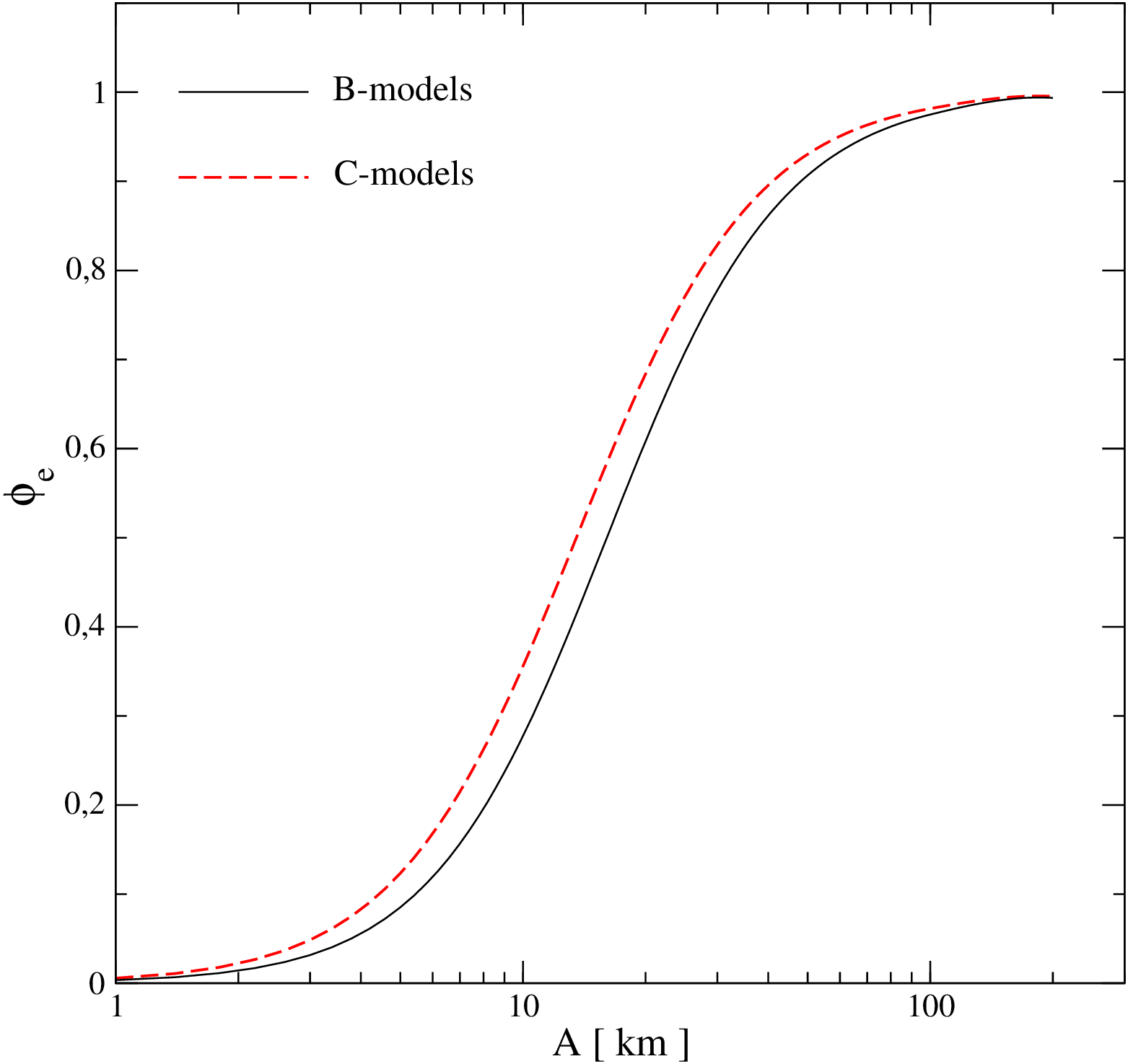}
\caption{This figure displays the dependence of $\phi_e = \Omega_e
/ \Omega_c$ on the parameter $A$ for the B (solid line)
and C (dashed line) models.
\label{fig:phi_A}}
\end{center}
\end{figure}

\begin{table}[h]
\begin{center}
\begin{ruledtabular}
\begin{tabular}{*{6}{c}}
  Model  &  $T_c$ & $\Om_c$  &   $\Om_e$  & $\varepsilon_e$ &  $J$ \\
$\left(A = 12~\textrm{km}\right)$   &  (ms)    & $\left(\times 10 ^{-2} ~{\rm km}^{-1}\right)$ &
                         $\left(\times 10^{-2}~{\rm km}^{-1}\right)$   & & $ \left( \rm{km}^2 \right)$  \\
\hline
  B1  &  1.719  & 1.218 &  0.435  & 0.161  & 0.935 \\
  B3  &  0.970  & 2.160 &  0.771  & 0.286  & 1.657 \\
  B6  &  0.657  & 3.189 &  1.139  & 0.422  & 2.447\\
\hline
  C1  &  1.719  & 1.218 &  0.535  &  0.129  & 0.832 \\
  C3  &  0.970  & 2.160 &  0.949  &  0.229  & 1.474 \\
  C6  &  0.657  & 3.189 &  1.401  &  0.338  & 2.176  \\
\end{tabular}
\end{ruledtabular}
\caption{\label{tab:RotModels} Parameters describing rotating models
of the B and C sequences, see also Table~\ref{tab:0Models}. Here $T_c$
and $\Om_c$ are respectively the period and the angular velocity on
the rotational axis, while $\Om_e$ and $\varepsilon_e$ represent the
angular velocity and the dimensioless parameter, described in
Eq.~(\ref{eq:eps_s}), at the equator. The angular momentum of the star
is denoted with $J$.}
\end{center}
\end{table}

\begin{table}[h]
\begin{center}
\begin{ruledtabular}
\begin{tabular}{*{6}{c}}
  Model  BJ1&  $T_c$ & $\Om_c$  &   $\Om_e$  & $\varepsilon_e$ &  $J$ \\
  A (km)    &  (ms)    & $\left(\times 10 ^{-2} ~{\rm km}^{-1}\right)$ &
                         $\left(\times 10^{-2}~{\rm km}^{-1}\right)$   & & $ \left( \rm{km}^2 \right)$  \\
\hline
    5  &  0.667  & 3.143  &  0.268  & 0.099  & 0.935 \\
    15 &  1.989  &  1.054 &  0.491  & 0.182  & 0.935 \\
    50 &  2.746  & 0.763  &  0.692  & 0.256  & 0.935\\
   100 &  2.834  &  0.739 &  0.725  &  0.268 & 0.935\\
\end{tabular}
\end{ruledtabular}
\caption{\label{tab:BJ1} This table displays the main properties of
the differentially rotating models BJ1  for four typical values of
$A$. }
\end{center}
\end{table}

\begin{table}[t]
\begin{center}
\begin{ruledtabular}
\begin{tabular}{c     c |     c   c  c  c  }
   & m & $\ell_{\rm max}=2$
         & $\ell_{\rm max}=3$ & $\ell_{\rm max}=4$ & $\ell_{\rm max}=5$ \\
\hline \hline
           &  -2      & 1.4793  & 1.4766   & 1.4768  & 1.4769    \\
${}^2 \textrm{f}$   &  -1      & 1.7037  & 1.7016   & 1.7016  &           \\
           &   1      & 2.1271  & 2.1252   & 2.1246  &           \\
           &   2      & 2.2932  & 2.2901   & 2.2913  & 2.2913    \\
\hline \hline
           &  -2      & 3.5798  & 3.5811   & 3.5815  & 3.5815     \\
${}^2 \textrm{p}_1$ &  -1      & 3.8156  & 3.8205   & 3.8206  &           \\
           &   1      & 4.4221  & 4.4278   & 4.4273  &           \\
           &   2      & 4.6467  & 4.6485   & 4.6488  & 4.6488    \\
\end{tabular}
\end{ruledtabular}
\caption{\label{tab:lmax} Frequencies ($\sigma/2\pi$), in kHz, of the
fundamental (${}^2 \textrm{f}$) and the first pressure mode (${}^2 \textrm{p}_1$) 
for the B1 stellar model with $A=12$~km and for $\ell_{\rm min} \le \ell_{\rm max} \le
\ell_{\rm min} + 3$.  }
\end{center}
\end{table}

\section{Numerical Results}
\label{Sec:Num.Results}

In this section, we solve Eqs.~(\ref{eq:u}-\ref{eq:xi}) and
investigate several aspects of the oscillation spectrum of
non-axisymmetric oscillations of slowly and differentially rotating
stars. In the first subsection we describe the procedure in
constructing the background stellar models while in the second
subsection the spectrum of non-axisymmetric oscillations is discussed
and we provide tables with frequencies for several background
configurations.
\begin{figure}[t]
\begin{center}
\includegraphics[width=85mm]{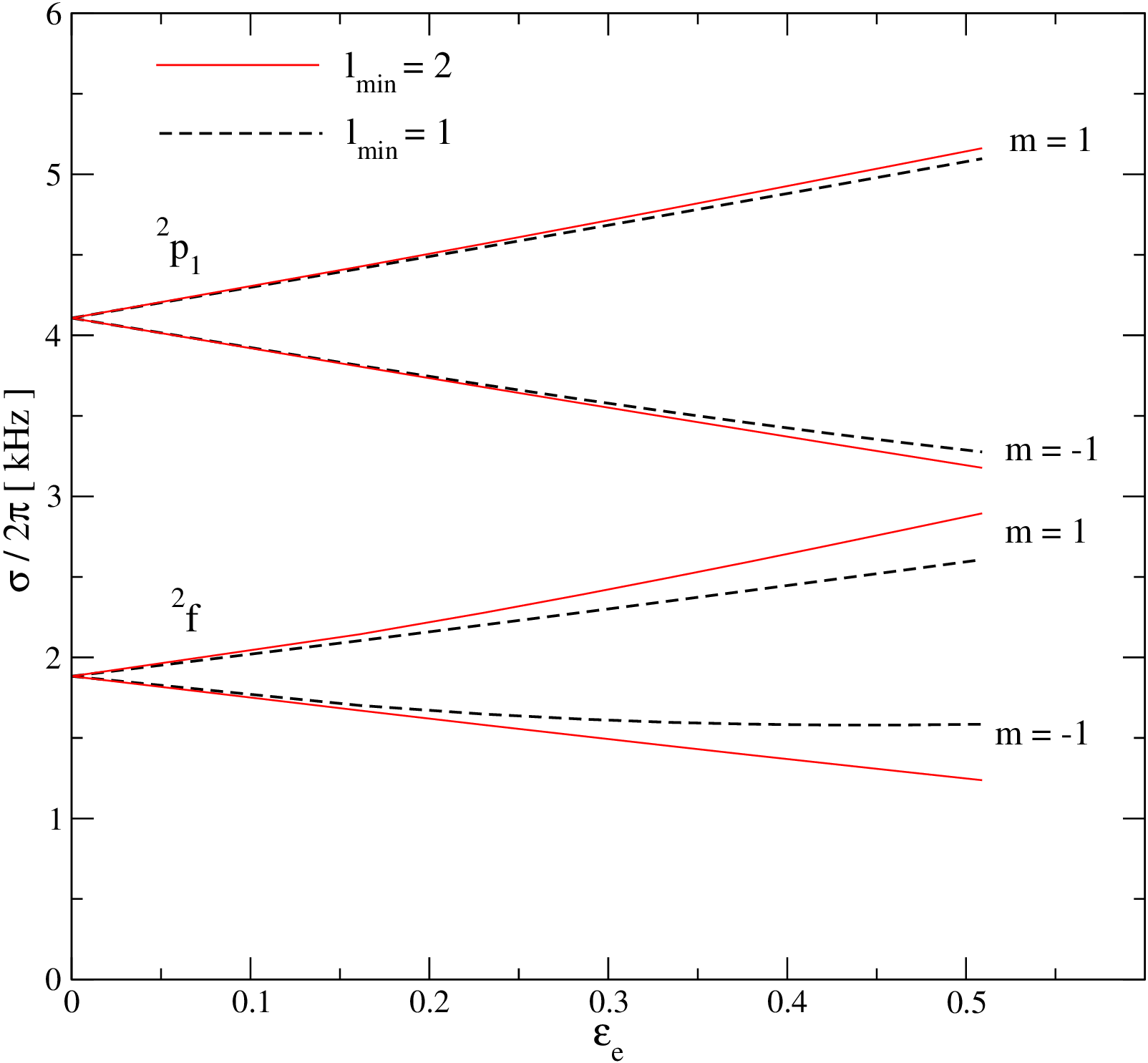}
\caption{Frequencies ($\sigma/2\pi$), in kHz, of the ${}^2 \textrm{f}$
 and ${}^2 \textrm{p}_1$ nonradial modes with $m=1$ for the B sequence
 of models having $A=12$~km.  The solid line corresponds to the purely
 first order slow rotation approximation, while the dashed line
 includes some second order corrections due to the component
 $u_{3}^{\ell m}$ of the perturbed fluid velocity.
 \label{fig:comp}}
\end{center}
\end{figure}

\subsection{Stellar Models}
\label{sec:Stmodels}

In the slow rotation approximation the stellar models maintain their
spherical shape and the rotation is treated as a small
perturbation. This approximation is practically applicable to all
known neutron stars even to those with rotation frequency of the a few
hundred Hz.  Here we specify a specific value for the central density
for a given EoS and we generated a sequence of rotating models by
varying the angular velocity of the star. For simplicity, we adopted
the relativistic barotropic EoS:
\begin{eqnarray}
p = K \rho ^{\Gamma} \, ,\qquad \qquad \eps = \rho + \frac{p}{\Gamma
-1} \, , \label{eq:EoS}
\end{eqnarray}
where $\rho$ is the rest mass density and $\eps$ the total energy
density, while $K$ and $\Gamma$ are the polytropic parameters.  We
have chosen the so called B-models ~\cite{Dimmelmeier:2005zk,
2007PhRvD..75f4019S}, which represent differentially rotating
polytropic stars with central rest mass density $\rho_c = 7.91 \times
10^{14}~\textrm{g cm}^{-3}$ and polytropic parameters $\Gamma=2$ and
$K=217.86~\textrm{km}^2$.  The nonrotating member of this sequence,
which is denoted as B0, has the typical neutron star mass
$M=1.4~M_{\odot}$ and radius $R=14.151$~km.  In order to investigate
the dependence of the non-axisymmetric spectra on the compactness of
the star, we constructed, in addition, a sequence of polytropic
stellar models with $0.102 \le M/R \le 0.216$.  The more compact model
$\left(M/R = 0.216\right)$ of this sequence will be called C-model
with $\rho_c = 2.0 \times 10^{15}~\textrm{g cm}^{-3}$.  More details
of the nonrotating members of the B and C sequences are provided in
Table~\ref{tab:0Models}.

In a differentially rotating star, the angular velocity on the
rotation axis $\Omega_c$ and the parameter $A$, which describes the
degree of differential rotation, are the other two free parameters
which need to be specified in constructing a sequence of rotating
stellar model by using Eq.~(\ref{drag-eq}). The angular velocity at
surface, $\Omega_s$, is then related to the one at the axis by the
following relation:
\begin{equation}
\Omega_s = \Omega_c \, \phi(A,\theta) \, ,
\end{equation}
where $\phi=\phi(A,\theta)$ is a scalar function that depends on the
law that describes the differential rotation. For the relativistic
j-constant rotation law given by Eq. (\ref{j-cons}) this function
reads:
\begin{equation}
\phi = \frac{1}{A^2 + e^{-2\nu} r^2 \sin^2\theta} \left( A^2 +
e^{-2\nu} r^2 \sin^2\theta \frac{\omega(r,\theta)}{\Omega_c} \right)
\,.
\end{equation}
The form of $\phi_e$ on the equator is drawn in Fig.~\ref{fig:phi_A}
for both B and C sequence of stellar models.  These two curves, for
practical reasons, can be Pad\'e approximated by the following
rational function:
\begin{equation}
\phi_e^{fit} = \frac{a_0 + a_1 A + a_2 A^2}{ 1 + a_3 A + a_4 A^2} \, ,
\label{eq:Pade}
\end{equation}
where the coefficients $a_k$ are listed in Table~\ref{tab:fitphi}.

The parameter that is commonly used in describing differentially
rotating stars is the ratio $T/|W|$, where $T$ is the rotational
kinetic energy and and $W$ the gravitational potential energy
respectively~\cite{Tassoul:1978}. Another possible rotational
parameter, which have been used in~\cite{1989MNRAS.239..153K}, is a
function of the total angular momentum, the total mass and the two
parameters defining polytrope.  In the relativistic slow rotation
approximation, the gravitational potential energy $W$ can be
accurately determined only at order $\mathcal{O} \left( \Omega ^2
\right)$, where the monopole and quadrupole corrections of the
gravitational mass and internal energy can be defined. Therefore, we
define as dimensionless rotation parameter the following ratio:
\begin{equation}
\veps_e \equiv \frac{\Omega_e}{\Omega_K} = \frac{\Omega_c}{\Omega_K}
\, \phi(A,\frac{\pi}{2}) = \veps_c \, \phi(A,\frac{\pi}{2}) \, ,
\label{eq:eps_s}
\end{equation}
where $\Omega_e$ represents the angular rotation at the stellar
equator and $\Omega_K$ is the Kepler angular velocity that defines the
mass shedding limit of a rotating star. In slowly rotating neutron
stars (first order in $\Omega$), $\Omega_K$ can be approximately
described by the angular velocity of a particle in a stable circular
Keplerian orbit at the equator of a nonrotating star $\Omega_K =
\sqrt{M/R^3}$.  Finally, the quantity $\veps_c \equiv
\Omega_c/\Omega_K$ is the value of the dimensionless rotation
parameter at the axis.  The parameters of a few selected models,
belonging to the B and C sequences, are shown in
Table~\ref{tab:RotModels} for $A=12~\textrm{km}$.
Differential rotation is described by a number of parameters
i.e. $\Omega_c$, $J$ and $A$. The effect of these parameters on the
oscillation spectrum will be studied for two families of background
stellar models.  In each family we fix one of the first two parameters
and vary $A$.  The first family of stellar models is the so called
B-sequence in which we fix the angular velocity $\Omega_c$ and we vary
the parameter $A$.  The properties of the various members of this
family, e.g. their angular velocity $\Om_e$ at the equator and of the
dimensionless parameter $\varepsilon_e$, can be estimated from those
of the $A=12~\textrm{km}$ model (Table~\ref{tab:RotModels}) by using
the Pad\'e expression~(\ref{eq:Pade}).
The second sequence is constructed by keeping constant the angular
momentum $J$ in the B-sequence of models and by varying the parameter
$A$.  The models of this new family will be named BJ models. The
angular momentum of the differentially rotating star can be determined
by the following expression~\cite{Hartle:1970ha}:
\begin{equation}
J = 2\pi \int_{0}^{R} \int_{0}^{\pi} d r ~ d \theta \left( \epsilon +
p \right) e^{\la-\nu} ~ \varpi ~ r^4 \sin^3 \theta \,
. \label{eq:Jdef}
\end{equation}
After expanding the variable $\varpi = \Omega - \omega$ in spherical harmonics
up to index $\ell=3$ and performing the angular integration,
Eq.~(\ref{eq:Jdef}) becomes:
\begin{equation}
J = \frac{8 \pi}{3} \int_{0}^{R} d r \left( \epsilon + p \right) e^{\la-\nu} ~
\varpi_1 ~ r^4 \, ,  \label{eq:Jfin}
\end{equation}
where only the $\ell=1$ component of $\varpi$ contributes to the
integral. In Table~\ref{tab:BJ1}, the main parameters characterizing
the BJ1 model are shown for selected values of $A$.

Once the sequence of BJ1 model is known, the rotational velocity
$\Omega_c^{\rm{BJn}}$ of another BJn model, with the same value of
$A$, can be estimated via the following relation:
\begin{equation}
 \Omega_c ^{~\rm{BJn}} = 
  \Omega_c ^{~\rm{BJ1}} ~ \frac{J^{\rm{Bn}}}{J^{\rm{B1}}} \, , 
\end{equation}
where $J^{\rm{B1}}$ and $J^{\rm{Bn}}$ refer to the angular momentum of
the B1 and Bn models respectively.

\begin{table}[!t]
\begin{center}
\begin{ruledtabular}
\begin{tabular}{c     c |     c   c  c  c  }
$A=12~\textrm{km}$ & & B0 & B1 & B3 & B6 \\ Modes & m & & & & \\
  \hline \hline & -2 & 1.883 & 1.479 & 1.171 & 0.839 \\ ${}^2
  \textrm{f}$ & -1 & 1.883 & 1.672 & 1.515 & 1.349 \\ & 1 & 1.883 &
  2.102 & 2.276 & 2.473 \\ & 2 & 1.883 & 2.293 & 2.614 & 2.970 \\
  \hline \hline & -2 & 4.107 & 3.579 & 3.177 & 2.744 \\ ${}^2
  \textrm{p}_1$ & -1 & 4.107 & 3.808 & 3.579 & 3.334 \\ & 1 & 4.107 &
  4.416 & 4.657 & 4.925 \\ & 2 & 4.107 & 4.647 & 5.068 & 5.534 \\
\end{tabular}
\end{ruledtabular}
\caption{\label{tab:f2a-mode} Frequencies ($\sigma/2\pi$), in kHz, of
the ${}^2 \textrm{f}$ and ${}^2 \textrm{p}_1$ nonradial modes for
selected B stellar models with $A=12$~km.}
\end{center}
\end{table}
\subsection{Spectral Properties}
\label{sec:spectrum}

Any nonradial mode of a rotating star is characterized by its harmonic
indices $(\ell, m)$, where $-\ell \le m \le \ell$. The axisymmetric
modes $m=0$ have been already discussed in Paper I, so here we focus
on the non-axisymmetric oscillations ($m\ne0$).

The non-axisymmetric modes of a rotating star split in corotating and
counterotating branches, whose pattern speed $\sigma_p = \sigma/m$ is
respectively positive and negative. Note that the modes are assumed to
behave as $e^{-\II \left( \sigma t - m \phi \right)}$.  In the slow
rotation approximation, the eigenfrequencies of any $(\ell,m)$
non-axisymmetric mode are linear functions of the rotation parameter
$\veps_e$ or $\veps_c$.  Note that $\varepsilon_c$ and $\varepsilon_e$
are related through Eq.~(\ref{eq:eps_s}).  For a given stellar model
the splitting of the modes, due to differential rotation, can be
described by the following relation:
\begin{equation}
\sigma^{\ell m} = \sigma_0^{\ell m} \pm \alpha(\ell,|m|,A) \, \veps_c =
\sigma_0^{\ell m} \pm \tilde{\alpha}(\ell,|m|,A) \, \veps_e\, ,
\label{eq:fre-lin}
\end{equation}
where $\alpha$ and $\tilde \alpha$ are scalar functions depending on
the harmonic indices and the differential parameter $A$ while
$\sigma_0$ is the oscillation frequency of the nonrotating
configuration with respect to an inertial observer.  For rotational
laws that are independent from the $\theta$ coordinate (e.g. uniform
rotation), relation~(\ref{eq:fre-lin}) is simplified and becomes
linear with respect to the azimuthal index $m$,
\begin{equation}
\sigma^{\ell m} = \sigma_0^{\ell m} + m \, \alpha^{\ell}_{\rm{uni}} \,
\veps_c \, ,
\end{equation}
as it is known from the Newtonian theory~\cite{1989nos..book.....U}.

\begin{figure}[t]
\begin{center}
\includegraphics[width=85mm]{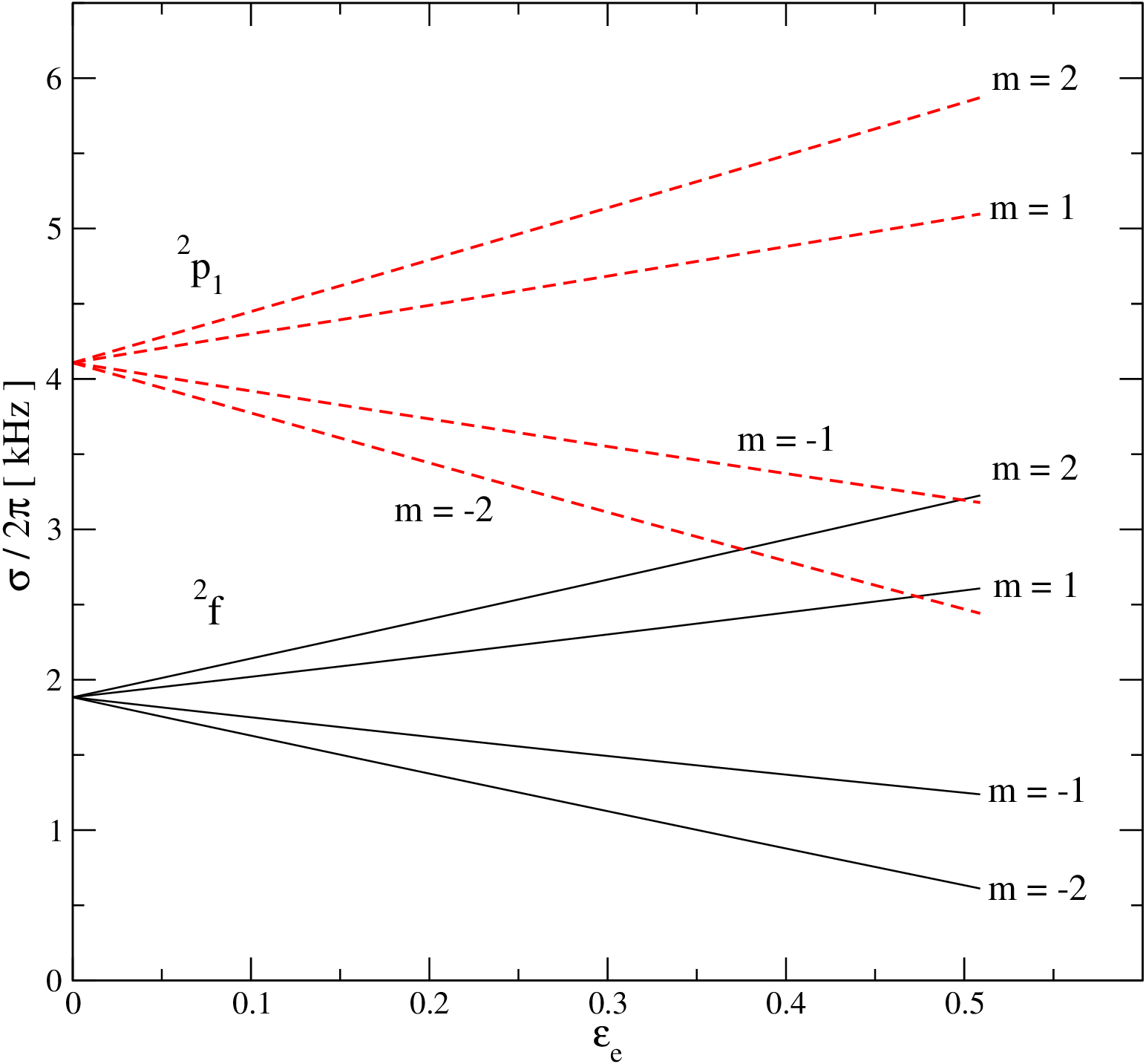}
\caption{ Frequencies ($\sigma/2\pi$), in kHz, of the ${}^2
\textrm{f}$ and ${}^2 \textrm{p}_1$ nonradial modes for B stellar
models with $A=12$~km.
\label{fig:Ba12-modes}}
\end{center}
\end{figure}
\subsubsection{Dependence on $\ell_{\rm max}$}

For any $(\ell,m)$ nonradial mode, Eqs.~(\ref{eq:u}-\ref{eq:xi}) can
be integrated by fixing the parameters $\ell_{\rm min}$ and $\ell_{\rm
max}$. The value of $\ell_{\rm min}$ is fixed by the selection of
index $m$ ($\ell_{\rm min}=|m|$), whereas the value of $\ell_{\rm
max}$ is chosen according to the maximum number of desired couplings
$n_c$ and it's upper limit is $\infty$.
Before proceeding in deriving any result it is important to test the
dependence of the oscillation frequencies for a given value of $\ell$
on $\ell_{\rm max}$.  In Table~\ref{tab:lmax} we present the results
of such a study, we actually show how the values of the fundamental
${}^2 \textrm{f}$ and pressure ${}^2 \textrm{p}_1$ modes of the B1
model depend on $\ell_{\rm max}$. The variation of $\ell_{\rm max}$
has not affected the actual values of the modes by more than
$0.2\%-0.3\%$. The same order of dependence of the mode frequencies on
$\ell_{\rm max}$ has been also observed for the other models of the B
and C-sequences.  Therefore, $\ell=2$ mode frequencies can be
accurately estimated by ignoring couplings with higher $\ell$'s
i.e. by keeping $\ell_{\rm max} = 2$.

In paper I, we discussed the role of the coupling between the polar
and axial perturbation functions in the axisymmetric case. Actually,
the component of the velocity perturbation $u_{3}^{\ell m}$ is
intrinsically of $\mathcal{O} \left( \Omega\right)$ order, (see
Eq.~(\ref{Eq_u3})). This implies that there are ``hidden'' second
order in $\Omega$ terms which induce some extra couplings into
Eqs.~(\ref{Eq_u1},\ref{Eq_u2}) that can influence the results. The
same type of problem is also present for the non-axisymmetric
perturbation, and will especially influence specific modes as for
example the $(\ell,m)=(2,1)$.
Thus we studied the $(\ell,m)=(2,1)$ eigenfrequencies either by using
the full coupled system of equations, i.e. $(\ell_{\rm min},\ell_{\rm
max})=(1,2)$, or by removing the ``hidden'' second order couplings,
i.e. $(\ell_{\rm min},\ell_{\rm max})=(2,2)$. The results are shown in
Fig.~\ref{fig:comp} and the similarities to the axisymmetric results
of Paper I are obvious.
That is, as we expected the dependence of the mode frequencies on the
rotation rate is not any more linear due to the presence of these
implicit second order rotational terms.  This effect is actually more
pronounced for the ${}^2 \textrm{f}$ mode.

We have chosen to neglect all of these ``hidden" second order terms in
order to get the expected linear relation between the oscillation and
the rotational frequencies.
\begin{figure}[t]
\begin{center}
\includegraphics[width=85mm]{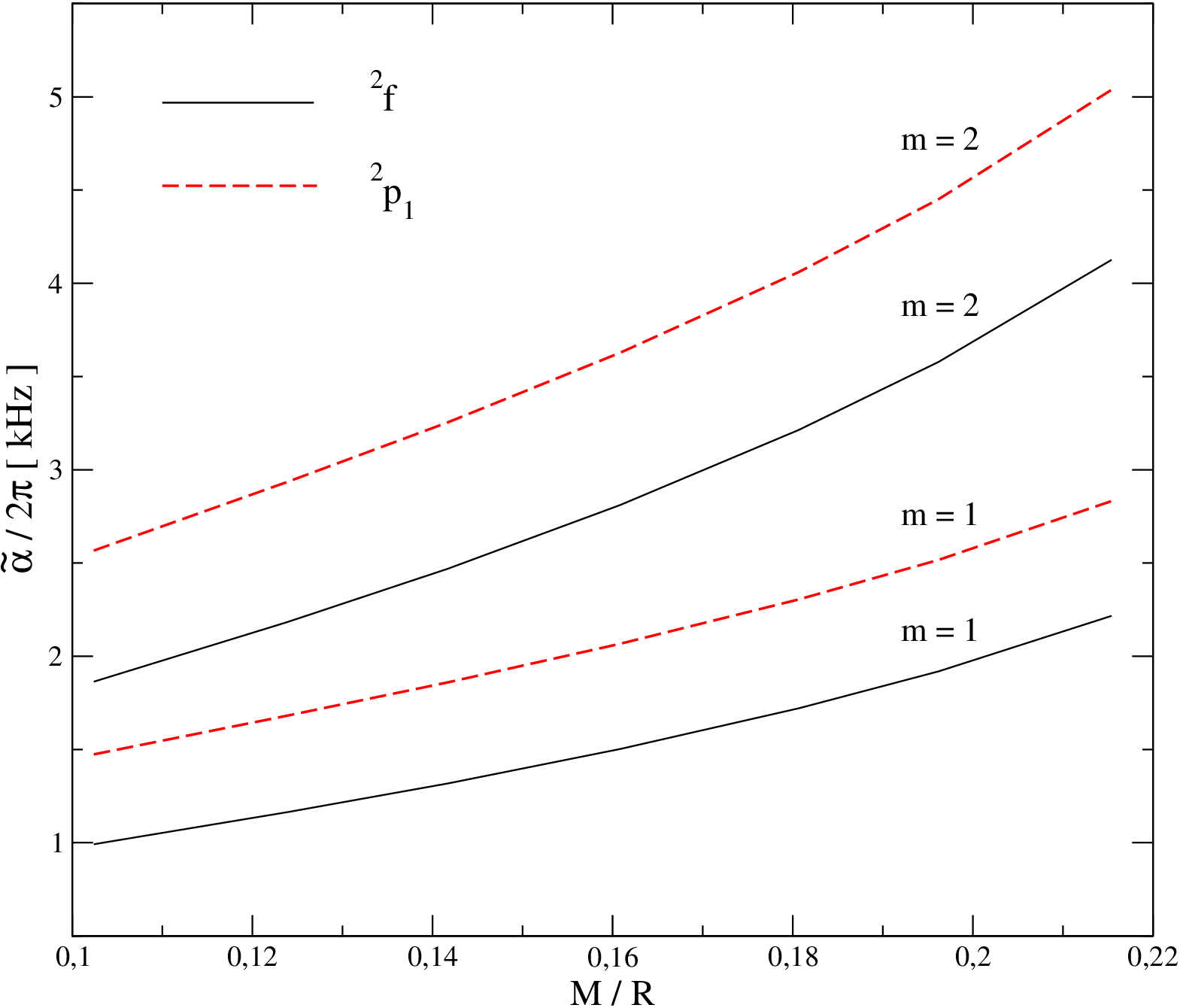}
\caption{ The dependence of the splitting factor ${\tilde \alpha}$ on
the stellar compactness is drawn both for the $^2 \textrm{f}$ and the
$^2 \textrm{p}_1$ modes.  Each member of this sequence of polytropic
stars has different compactness but the same $\Omega_e/\Omega_c$
ratio.
 \label{fig:f2-b-c-models}}
\end{center}
\end{figure}
\subsubsection{Eigenfrequencies}

Here, we study how the mode frequencies depend on the rotation
parameter $\varepsilon_e$, the compactness $M/R$ of the star and the
degree of differential rotation $A$.

Let us start by choosing the B sequence of stellar models for a fixed
value of $A=12$~km, these are the equilibrium configurations already
used in~\cite{Dimmelmeier:2005zk, Stergioulas:2003ep}.  Due to
rotation, these modes are split in two symmetrical branches
characterized by the azimuthal number $m$.  In
Fig.~\ref{fig:Ba12-modes} the splitting induced by rotation for the
$\ell = 2$ fundamental and first pressure modes is shown while the
actual values of the eigenfrequencies for these sequence of models are
given in Table~\ref{tab:f2a-mode}.

Stellar compactness affects significantly the mode splitting, but one
should be careful on how to quantify this effect. In fact, the angular
velocity profile of the stellar model depends on the compactness and
it is inconsistent to create a sequence of models by just keeping $A$
constant. Instead, we have choosen to compare stellar models according
to the ratio between the angular velocity at the equator and at the
rotational axis, i.e. by keeping $\phi_e = \Omega_e/\Omega_c$
constant.  We then consider the B sequence of stellar models with
$A=12$~km as reference for generating other polytropic stars described
by the Eq.~(\ref{eq:EoS}).  All these models have $\phi_e = 0.3572$,
which for the C models corresponds to $A=10.08$~km.  In
Figure~\ref{fig:f2-b-c-models} we show the dependence of the
``splitting" coefficient $\tilde{\alpha}$~(\ref{eq:fre-lin}) on the
stellar compactness for the ${}^2 \textrm{f}$ and ${}^2 \textrm{p}_1$
modes.  It is obvious that the splitting is enhanced by the
compactness and it can be easily be twice as large for very compact
neutron stars. In principle, within the gravitational wave
asteroseismology \cite{2001MNRAS.320...307K}, this effect of rotation
can be used, to infer the rotation and compactness of the oscillating
neutron star.

\begin{figure}[t]
\begin{center}
\includegraphics[width=85mm]{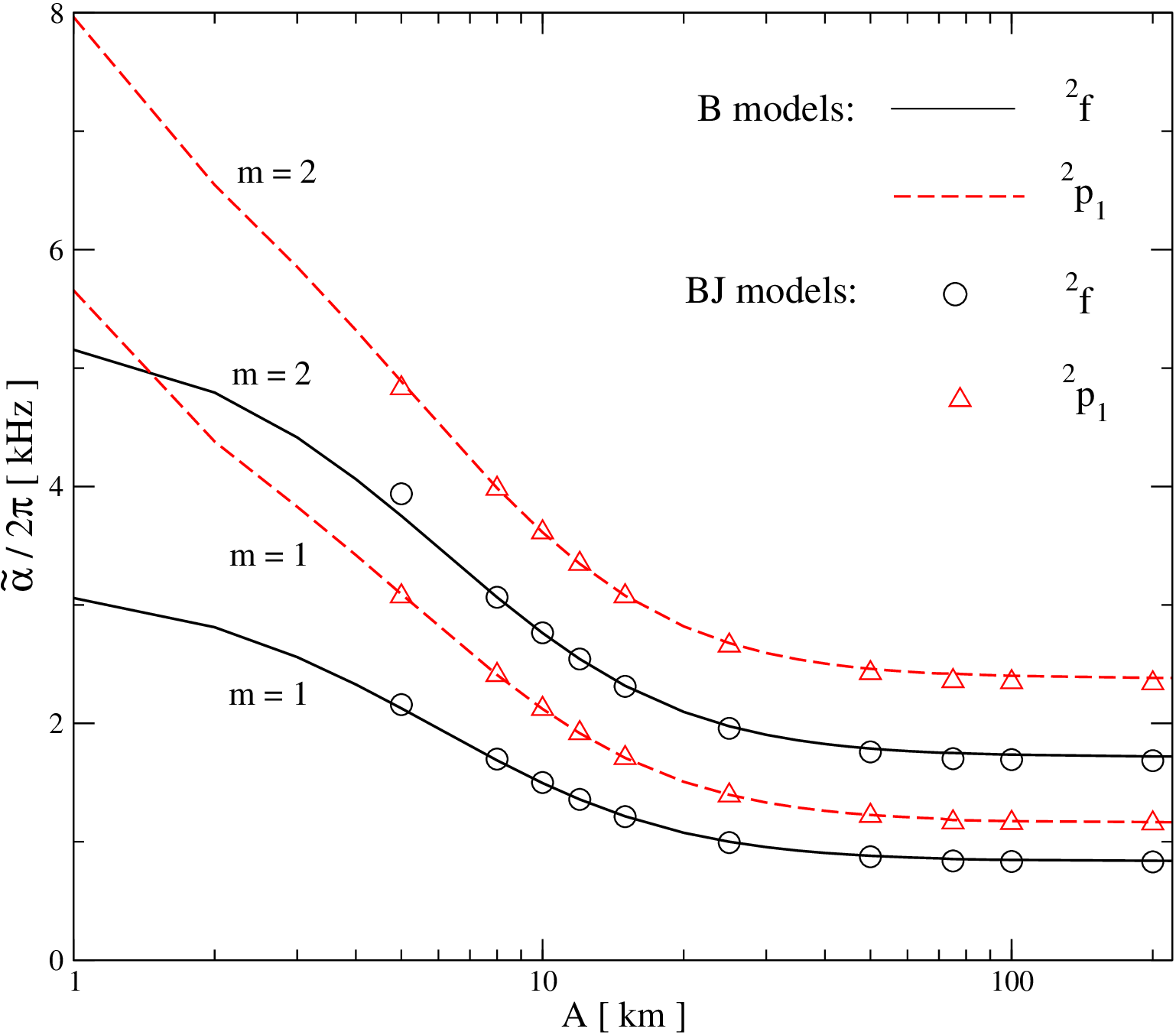} 
\caption{The dependence of the splitting factor ${\tilde \alpha}$ on
the differential rotation parameter $A$ for the B and BJ
models is drawn for the $^2 \textrm{f}$ and $^2 \textrm{p}_1$
modes. The solid and the dashes lines represents respectively the
${}^2f$ and ${}^2p_{1}$ modes of the B models, whereas the
open circles and triangles the ${}^2f$ and ${}^2p_{1}$ modes of the BJ
models.
\label{fig:f2-B-a12-100}}
\end{center}
\end{figure}

Finally, we investigate the dependence of the non-axisymmetric modes
on the parameter $A$ describing the degree of differential rotation.
We consider the two sequences of stellar models described earlier in
Sec.~\ref{sec:Stmodels}. These are, the B sequence of stellar models
with $\Omega_c$ constant and the BJ sequence with $J$ constant.  In
Table~\ref{tab:f2-mode}, we report the frequencies of the
${}^2\textrm{f}$ and ${}^2 \textrm{p}_1$ modes for some of these
models.  In Fig.~\ref{fig:f2-B-a12-100}, it is shown as the value of
$A$ affects the splitting of the ${}^2 \textrm{f}$ and ${}^2
\textrm{p}_1$ modes.  The splitting of the eigenfrequencies for
stellar models with high degree of differential rotation depends
strongly on the rotation parameter $\varepsilon_e$. Furthermore, it is
worth noticing that the B and BJ models show the same dependence of
$\tilde \alpha$ with respect to $A$. This behavior is expected since
the differential rotation is described in both stellar sequences by
the relativistic j-constant rotation law~(\ref{j-cons}) and the
no-rotating model of the sequences is the same.

As it has been mentioned in the introduction, differential rotation
plays an important role on the onset of rotational
instabilities. Non-axisymmetric pulsations of rotating stars can
become dynamical or secular unstable and enhance gravitational wave
emission, e.g. see~\cite{Andersson:2002ch,lrr-2003-3} and references
there in.  These instabilities appear when the rotational parameter
$\beta = T/|W|$, reaches some critical value $\beta_c$. The onset of
secular and dynamical bar mode instabilities in Newtonian
incompressible and uniformly rotating bodies is at $\beta_c = 0.14$
and $\beta_c = 0.27$ respectively. The secular instabilities are
driven by dissipative processes, such as gravitational radiation via
the Chandrasekhar-Friedman-Schutz (CFS) mechanism and viscosity.

\begin{figure}[t]
\begin{center}
\includegraphics[width=85mm]{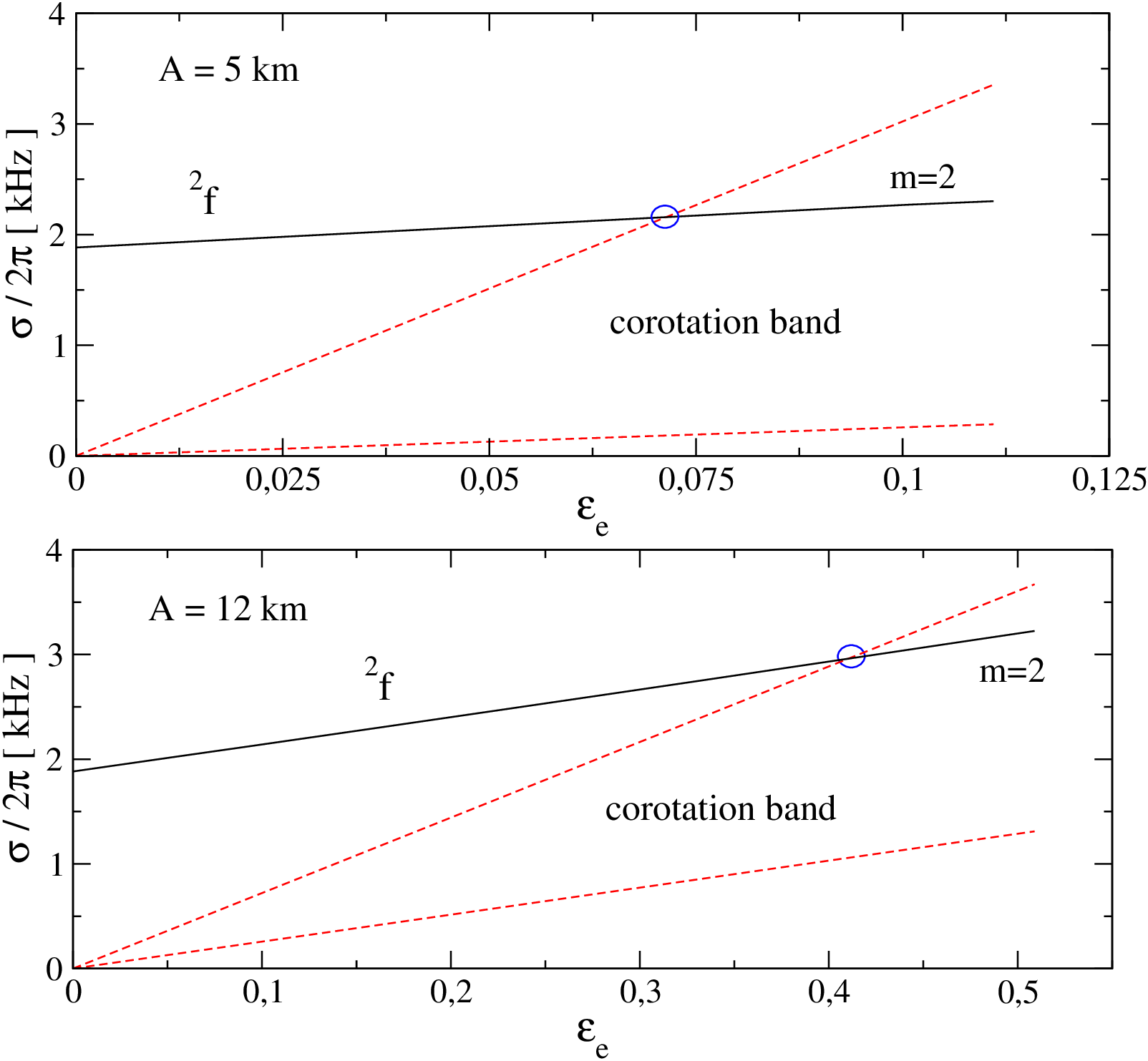}
\caption{The upper panel displays the $\ell=m=2$ f-mode (solid line)
for the B sequence of stellar models with a high degree of
differential rotation $A=5$~km.  The limiting lines of the corotation
band are represented in dashed lines.  The lower panel is similar but
for $A=12$~km.
\label{fig:corr-band}}
\end{center}
\end{figure}

\begin{figure}[t!]
\begin{center}
\includegraphics[width=85mm]{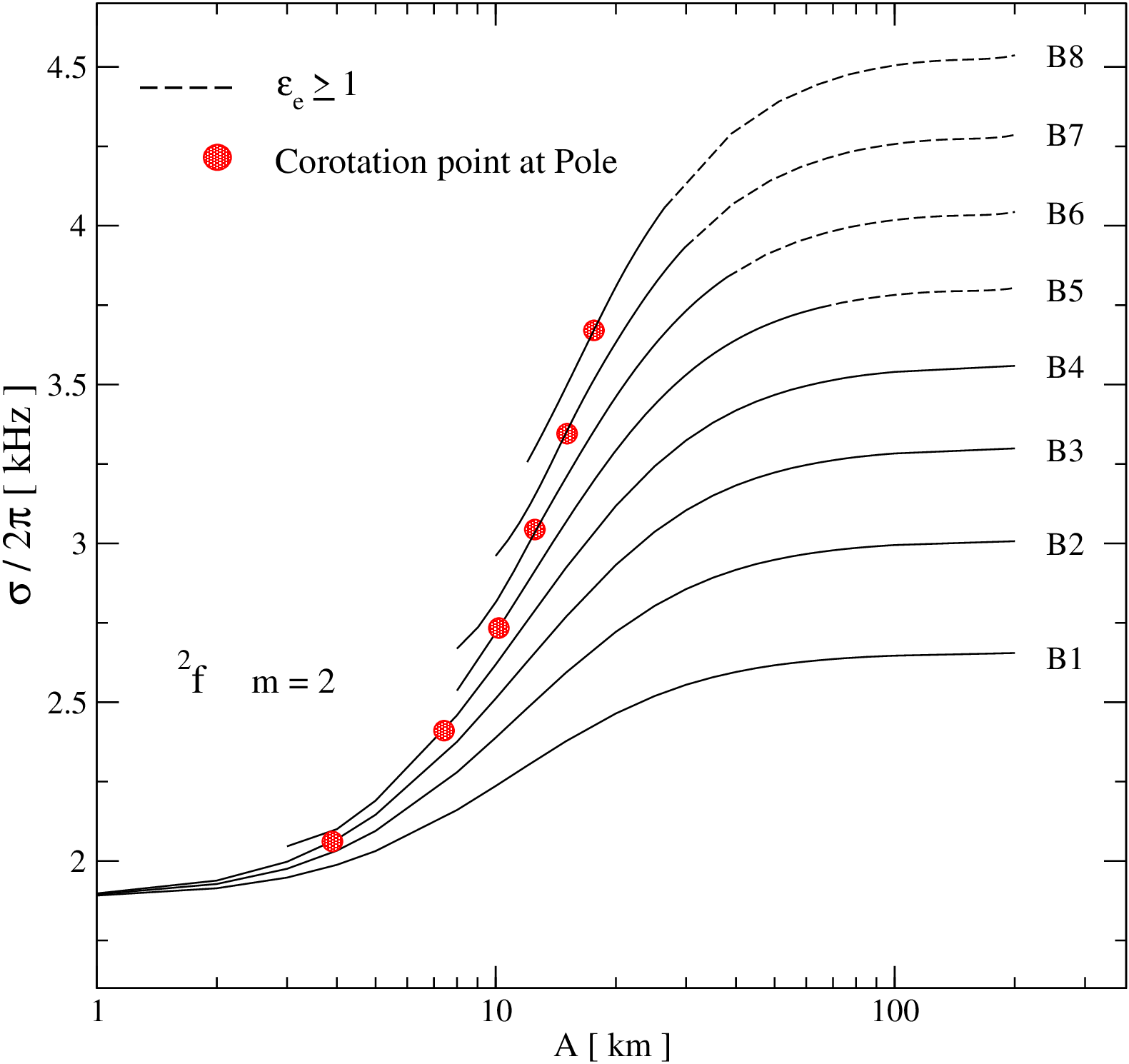}
\caption{For B stellar models, the figure displays the variation of
the $\ell=m=2$ fundamental mode with respect to the differential
parameter $A$. In any sequence, the circles denotes when the
fundamental mode goes in corotation. For $A$ lower then these values,
the fundamental mode is always in corotation. The dashed line denotes
instead the stellar models that are rotating faster than the mass
shedding limit.
\label{fig:f2-B-A}}
\end{center}
\end{figure}

 Actually, the so called ``low $T/W$'' dynamical instability appears
in stellar models with high degree of differential rotation. Several
studies carried out with nonlinear hydrodynamical codes
\cite{2001ApJ...550L.193C,2006ApJ...651.1068O} and perturbative
methods \cite{2005ApJ...618L..37W,2006MNRAS.368.1429S} suggest a
strong correlation between the onset of the low $T/W$ instability and
the presence of corotation modes. By definition, these modes have
their pattern speed equal to the local angular velocity of the star,
i.e. $\sigma/m = \Omega(r,\theta)$.  For a differentially rotating
star the possible corotation band of the spectrum is given by the
interval $ \Omega_e \le \sigma / m \le \Omega_c$, which is obviously
larger for highly differentially rotating stars, i.e. with smaller
$A$.  For the B sequence of stellar models, we estimated the required
differential and rotation parameters for having corotation
modes. Initially, we show in Fig.~\ref{fig:corr-band} the cases
$A=5$~km and $A=12$~km. As expected, for $A=5$~km the $\ell=m=2$
fundamental mode ``corotates'' for smaller rotation rate
$\varepsilon_e$ than the $A=12$~km configuration.
Fig.~\ref{fig:f2-B-A} shows the variation of the $\ell=m=2$
fundamental mode with respect to the parameter $A$.  All these models
with the exception of the B1 and B2 present a corotation ${}^2
\textrm{f}$ mode.  The upper limit of the parameter $A$ for having
corotation is indicated with $A_c$ and it is illustrated with a circle
in Fig.~\ref{fig:f2-B-A}.  Therefore, the star has corotation modes
when $A\le A_c$. In Table~\ref{tab:f2-Ac} the critical values $A_c$
and the corresponding eigenfrequencies are listed for various models
of the B sequence.  When non-axisymmetric modes are into the
corotation band the eigenvalue problem is mathematically
singular. Still, for some stellar models it was possible to isolate
the eigenmode frequencies by carefully studying their eigenfunctions
into the continuous spectrum~\cite{2006MNRAS.368.1429S}.

\begin{table}[!t]
\begin{center}
\begin{ruledtabular}
\begin{tabular}{c c |  c |  c  c c  | c c c   }
   Modes           & $m$ & A\,(km) & B1 & B3  & B6 & BJ1  & BJ3      & BJ6   \\ 
 \hline \hline     &  1  & 5       & 1.965   & 2.029 & 2.101  & 2.097    &  2.269   &  \\ 
 ${}^2 \textrm{f}$ & 1   & 12      & 2.102   & 2.276 & 2.473  & 2.102    &  2.276   & 2.473 \\
                   & 1   & 50      & 2.244   & 2.535 & 2.864$^{\ast}$  & 2.107      &  2.285   & 2.484 \\ 
                   & 1   & 100     & 2.256   & 2.101 & 2.896$^{\ast}$  & 2.107      &  2.285   & 2.483 \\ 
 \hline \hline     & 2   & 5       & 2.027   & 2.138 & 2.627  & 2.274  &            &           \\      
 ${}^2 \textrm{f}$ & 2   & 12      & 2.293   & 2.614 & 2.970  & 2.293  &  2.614     &           \\ 
                   & 2   & 50      & 2.615   & 3.220 & 3.917$^{\ast}$  & 2.334  &  2.698   & 3.113  \\ 
                   & 2   & 100     & 2.647   & 3.283 & 4.018$^{\ast}$  & 2.338  &  2.707   & 3.128  \\ 
\hline \hline      
                    & 1  & 5       & 4.226   & 4.317 & 4.416  & 4.412  &  4.648   &  4.908   \\
${}^2 \textrm{p}_1$ & 1  & 12      & 4.416   & 4.657 & 4.925  & 4.416  &  4.657   &  4.925    \\ 
                    & 1  & 50      & 4.609   & 5.011 & 5.465$^{\ast}$  & 4.419  &  4.665   &  4.940    \\ 
                    & 1  & 100     & 4.624   & 5.038 & 5.508$^{\ast}$  & 4.418  &  4.664   &  4.938     \\ 
\hline \hline 
                    & 2  & 5       & 4.295   & 4.438 & 4.593  & 4.586  &  4.949   &   5.486   \\ 
${}^2 \textrm{p}_1$ & 2  & 12      & 4.647   & 5.068 & 5.534  & 4.647  &  5.068   &   5.534    \\ 
                    & 2  & 50      & 5.114   & 5.947 & 6.912$^{\ast}$  & 4.728  &  5.229   &   5.799    \\ 
                    & 2  & 100     & 5.164   & 6.044 & 7.067$^{\ast}$  & 4.737  &  5.247   &   5.829    \\
\end{tabular}
\end{ruledtabular}
\caption{\label{tab:f2-mode} The mode frequencies ($\sigma/2\pi$) of
the fundamental (${}^2 \textrm{f}$) and the first pressure mode (${}^2
\textrm{p}_1$) in kHz, for the B and BJ sequences of stellar
models. The B6 models for $A> 38~\rm{km}$ are rotating faster than the
mass sheeding limit, their eigenmodes are labeled with a star. In the
BJ3 and BJ6 models, some of the eigenfrequencies cannot be determined
as they are inside the continuous spectrum. Therefore, we leave a
blank space in the table.}
\end{center}
\end{table}


\section{Conclusions and Discussion\label{conclusions}} \label{sec:concl}
We presented a first comprehensive study of non-axisymmetric
oscillations of slowly and differentially rotating neutron stars in
the perturbative framework of General Relativity. By using Cowling
approximation, we examined the spectral properties of polytropic stars
and investigated their dependence on four main parameters: stellar
compactness $M/R$, stellar rotation rate at the equator $\veps_e$,
degree of differential rotation $A$ and the maximum number of
perturbative couplings $\ell_{\rm max}$.  In accordance with the first
order slow rotation approximation, the non-axisymmetric modes exhibit
a linear splitting with respect to the rotational parameter $\veps_e$.
However, in some of the eigenmode patterns appears a quadratic
deviation with respect to the expected linear behavior. This is due to
the presence of ``implicit'' second order perturbative terms in the
perturbation equations. We have identified and then neglected these
terms in order to be consistent with the order of approximation
adopted for the background spacetime.  Moreover, we show that the
non-axisymmetric spectrum can be described by including only a small
number of couplings between perturbation functions.  For instance, we
determined the quadrupolar spectrum with an accuracy to better than
1\% by setting $\ell_{\rm max}=2$, which corresponds to the lowest
possible number of coupling terms in the perturbation equations.

We found that both differential rotation and stellar compactness
affect the non-axisymmetric spectrum. In fact, the rotational
splitting of the non-axisymmetric modes is enhanced by stellar
compactness and the degree of differential rotation.

For the study of the "low $T/W$" instability we calculated the
corotation band of some polytropic models and the necessary rotational
configuration for having a corotating quadrupolar fundamental mode.
Moreover, we verified the results of Newtonian studies, i.e. that
the value of the rotational parameter $\epsilon_e$, which is required for having a corotating
fundamental mode, is inversely proportional to the degree of
differential rotation.
The onset and the details of the low $T/|W|$ dynamical instability of
differentially rotating stars will be the subject of a future
investigation.

\[ \]
{\bf Acknowledgments:} We thank N. Stergioulas, H. Sotani
and M. Vavoulidis for helpful discussions.  A.P. is supported by a
``Virgo EGO Scientific Forum'' (VESF) and by the EU program
ILIAS. A.S. was supported by the Pythagoras II grand of the Greek
Ministry of Research and Development and by IKY post-doctoral grand of
the Ministy of Education and is grateful to the Groupe Gravitation 
Relativiste et Cosmologie (GR$\varepsilon$CO) of the Institut
d'Astrophysique de Paris for its hospitality while this work was being completed. 
This work is supported in part by the National Science Foundation under
grant number PHY 06-52448, and by the National Aeronautics
and Space Administration under grant number NNG06GI60G.
This work was also supported by the German Foundation (DFG) via SFB/TR7. 
We also thank the anonymous referee for useful suggestions that have
improved the structure and the presentation of our work.

\begin{table}[!t]
\begin{center}
\begin{ruledtabular}
\begin{tabular}{  c  c  c  c  }
  Models           &$\varepsilon_e$ &   $A_c$  &  $\sigma / 2\pi$  \\
                   &                &   $\left( {\rm km} \right)$
                   &   $\left( {\rm kHz} \right)$    \\
\hline \hline
   B3              & 0.042  &   3.898     & 2.0615     \\
   B4              & 0.162  &   7.422     & 2.4105     \\
   B5              & 0.302  &  10.182     & 2.7336     \\
   B6              & 0.446  &  12.534     & 3.0439     \\
   B7              & 0.610  &  15.095     & 3.3454     \\
   B8              & 0.778  &  17.625     & 3.6707      \\
 \end{tabular}
\end{ruledtabular}
\caption{\label{tab:f2-Ac} Critical values of the differential
rotational parameters for having corotation modes. These values
correspond to the $\ell=m=2$ f-mode.}
\end{center}
\end{table}
\appendix

\section{Coefficient Matrices \label{app:matrix}}

In Sec.~\ref{sec:pert-eqs}, we  wrote the fluid perturbations as
three infinite dimensional vectors $\U$, $\Sv$ and $\V$, see
Eq.~(\ref{def:USVvec}). These vectors can be written in terms of the
2-vectors $u^{\ell m}$, $s^{\ell m}$ and $v^{\ell m}$ that are defined
in Eq.~(\ref{def:usv}). They contain the fluid perturbations with
harmonic indices $(\ell,m)$.  These definitions enable us to describe
the perturbation equations in a more compact form by introducing a set
of infinite dimensional linear matrix operators,
i.e. $\mathcal{A}_{\sigma}^{\rm tot}, \mathcal{C}, \mathcal{D},
\mathcal{S}_{\sigma}, \mathcal{M}, \mathcal{N},
\mathcal{Q}_{\sigma}^{\rm tot}$.  In this Appendix, we provide the
transformation laws of these operators.  Actually, it is much more
convenient to define how these operators act on the single vectors
i.e. the ones defined in Eq. (\ref{def:usv}), rather than dealing with
the infinite dimensional vectors~$\U$, $\Sv$ and $\V$ given by
Eq. (\ref{def:USVvec}).  Let us start with the
operator~$\mathcal{A}_{\sigma}^{\rm tot}$ used in Eq.~(\ref{eq:u}), which
can be written as the sum of two operators $\mathbf{A}_{\sigma}$ and
$\mathbf{B}^{\pm 2}$ defined as follows:
\begin{equation}
\mathbf{A}_{\sigma} \equiv \mathbf{ \hat{A}}  \mathcal{I} \, , \qquad \qquad
\mathbf{B}^{\pm2}   \equiv \mathbf{ \hat{B}}  \mathcal{L}_1^{\pm 2 } \, ,
\end{equation}
where $\mathbf{ \hat{A}}$ is the following $2\times2$ matrix:
\begin{eqnarray}
 \mathbf{\hat A}  =
  \left(\begin{array}{cc}
    \left( \frac{\Gamma_1}{\Gamma} - 1 \right) \nu' c_s^{-2}     &  - \sigma + m  \left( \Omega_1 + 6 \Omega_3 \right) \\
\\
    \left[ \left( \sigma - m \left( \Omega_1 + 6 \Omega_3 \right) \right) \frac{1}{c_s^2}
    + 2 m \left( \varpi_1 +  6 \varpi_3 \right)
   \right]  e^{2 \left(\lambda -\nu \right)}
    &  - \frac{2}{r} + \lambda' - 2 \nu' + \frac{\nu'}{c_s^2}   \\
 \end{array}\right)\, ,
\end{eqnarray}
where $\mathcal{I}$ is the identity matrix  and $\mathcal{L}_1^{\pm 2 }$
is an operator defined in Eq.  (\ref{eq:L1pm2}).   The action of $\mathbf{A}_{\sigma}$  on
$u^{\ell m}$ is given by:
\begin{equation}
\mathbf{A}_{\sigma} u^{\ell m} \longmapsto \mathbf{ \hat{A}} u^{\ell m} \, ,
\end{equation}
whereas by the definition of $\mathcal{L}_1^{\pm 2 }$, the operator
$\mathbf{B}^{\pm 2}$ can transform $u^{\ell m}$ in three ways:
\begin{equation}
\mathbf{B}^{\pm 2} u^{\ell m} \longmapsto \mathbf{ \hat{B}}
 \left[ - Q_{\ell-1 m} Q_{\ell m} u^{\ell-2 m}, \left( 1
- Q_{\ell m}^{2} - Q_{\ell+1 m}^{2}\right) u^{\ell m}, -  Q_{\ell m} Q_{\ell+1 m} u^{\ell+2 m}
\right] \, ,
\end{equation}
where the coefficient $Q_{\ell m}$ is defined in Eq.~(\ref{eq:Qlm})
and $\mathbf{\hat B}$ is the following $2\times2$ matrix:
\begin{eqnarray}
 \mathbf{\hat B}  =
  \left(\begin{array}{cc}
    0      &  - \frac{15}{2} \, m \, \Omega_3  \\
    \frac{15}{2} \, m  \, \left( \frac{\Omega_3}{c^s_2}
   - 2 \varpi_3 \right) e^{ 2 \left( \lambda-\nu \right)}        &   0    \\
 \end{array}\right) \, .
\end{eqnarray}
Note that the operator $\mathbf{B}^{\pm 2}$ couple perturbations with
different index $\ell$ while $\mathbf{A}_{\sigma}$ does not.

The other two operators $\mathcal{C}$ and $\mathcal{D}$ of
Eq.~(\ref{eq:u}) are defined  as follows:
\begin{eqnarray}
\mathcal{C} &\equiv& \mathbf{\hat C}_{0} \mathcal{I} ~ \oplus ~
                     \mathbf{\hat C}_{1} \mathcal{L}_{1}^{\pm 1} ~ \oplus ~
                     \mathbf{\hat C}_{2} \mathcal{L}_{1}^{\pm 2} ~ \oplus ~
                     \mathbf{\hat C}_{3} \mathcal{L}_{1}^{\pm 3} \, , \nn \\
\mathcal{D} &\equiv& \mathbf{\hat D}_{0} \mathcal{I} \, ,
\end{eqnarray}
where $\mathcal{L}_{1}^{\pm 1}$, $\mathcal{L}_{1}^{\pm 2}$ and
$\mathcal{L}_{1}^{\pm 3}$ are given in Sec.~\ref{Integrals}, and the
matrices $\mathbf{\hat C}_{0}$, $\mathbf{\hat C}_{1}$, $\mathbf{\hat
C}_{2}$, $\mathbf{\hat C}_{3}$, $\mathbf{\hat D}_{0}$ have the following expressions:
\begin{eqnarray}
 \mathbf{\hat C}_{0}    =
  \left(\begin{array}{cc}
     - m \left( f_1 + 6 f_3 \right)              &   0   \\
     \ell \left( \ell +1 \right) e^{2\la} r^{-2}    &   0    \\
 \end{array}\right) \, ,  \qquad
 \mathbf{\hat C}_{1}  =
  \left(\begin{array}{cc}
    0      &  - \left( f_1 + 6 f_3 \right)   \\
    0      &   0    \\
 \end{array}\right) \, ,
\end{eqnarray}
\begin{eqnarray}
 \mathbf{\hat C}_{2}   =
  \left(\begin{array}{cc}
     \frac{15}{2} \, m \, f_3  &  0 \\
                         0    &   0    \\
 \end{array}\right)   \, ,  \qquad
 \mathbf{\hat C}_{3}  =
  \left(\begin{array}{cc}
    0      &  \frac{15}{2} \,  f_3    \\
    0      &   0    \\
 \end{array}\right) \, ,   \qquad
 \mathbf{D}  =
  \left(\begin{array}{cc}
    \nu' c_s^{-2}      &  0    \\
    0      &   0    \\
 \end{array}\right) \,.
\end{eqnarray}
and
\begin{eqnarray}
f_1  = 2 \left( \frac{1}{r} - \nu' \right) \varpi_1 - \omega_1 '  \, , \qquad   \qquad
f_3  = 2 \left( \frac{1}{r} - \nu' \right) \varpi_3 - \omega_3 '  \, .
\end{eqnarray}
The operation of $\mathcal{L}_{i}^{\pm j}$ for $j=1,2,3$ on $u^{\ell
m}$ follows the previous description as for the operator
$\mathbf{B}^{\pm 2}$.

In Eq.~(\ref{eq:s}) the operators $\mathcal{S_{\sigma}}$ and
$\mathcal{M}$ are defined as follows:
\begin{eqnarray}
\mathcal{S}_{\sigma} & = &  \mathbf{\hat \Sigma}_{\sigma} \mathcal{I} ~\oplus~ \mathbf{\hat K}_1 \mathcal{L}_{3}^{\pm 1}
                              ~\oplus ~\mathbf{\hat K}_2 \mathcal{L}_{4}^{\pm 1}
                              ~\oplus ~\mathbf{\hat J}_1 \mathcal{L}_{2}^{\pm 2}
                              ~\oplus ~\mathbf{\hat J}_2 \mathcal{L}_{3}^{\pm 2}
                              ~\oplus ~\mathbf{\hat J}_3 \mathcal{L}_{4}^{\pm 2}
                              ~\oplus ~\mathbf{\hat W}   \mathcal{L}_{2}^{\pm 3}  \, , \\
\mathcal{M}           & = &  \mathbf{\hat M}_0 \mathcal{I} ~ \!\oplus~ \mathbf{\hat M}_1 \mathcal{L}_{2}^{\pm 1}
                              ~\! \oplus~ \mathbf{\hat M}_2 \mathcal{L}_{1}^{\pm 2}
                              ~\oplus~ \mathbf{\hat M}_3 \mathcal{L}_{3}^{\pm 3}  \, ,
\end{eqnarray}
where the quantities with a hat are matrices. Those associated with
the operator $\mathcal{S_{\sigma}}$ have the following form:
\begin{equation}
    \mathbf{\hat \Sigma}_{\sigma}  =  \left[  - \sigma  + m \left( \Omega_1 + 6 \Omega_3 \right)
     -  \frac{2 m}{\Lambda} \left( \varpi_1 + 6 \varpi_3
     \right) \right] \mathbf{I}_{2\times2} \, , 
\end{equation}
\begin{eqnarray}
 \mathbf{\hat K}_1   = - \frac{2 \left( \varpi_1 + 6  \varpi_3 \right) }{\ell \left( \ell + 1 \right)}
  \left(\begin{array}{cc}
       0 & \quad 1   \\
       1 & \quad 0     \\
 \end{array}\right)\,,   \qquad  \qquad
 \mathbf{\hat K}_2  = \frac{15 m^2 }{\ell \left( \ell + 1 \right)}
  \left(\begin{array}{cc}
       0 &      2 \varpi_3 + \Omega_3     \\
     2  \varpi_3   & 0
\\
 \end{array}\right)\,,
\end{eqnarray}
\begin{eqnarray}
 \mathbf{\hat J }_1  = \frac{ 15  m}{\ell \left( \ell + 1 \right)}
  \left(\begin{array}{cc}
     \Omega_3 - 2 \omega_3        &    0 \\
    0  &       2 \varpi_3   \\
 \end{array}\right)\,,
\qquad \qquad
 \mathbf{\hat J}_2   = \frac{15  m}{\ell \left( \ell + 1 \right)}
  \left(\begin{array}{cc}
     2 \varpi_3    &    0      \\
     0             &   \Omega_3 - 2 \omega_3   \\
 \end{array}\right)\,,
\end{eqnarray}
\begin{eqnarray}
 \mathbf{\hat J}_3  = - \frac{15}{2} \frac{m}{\ell \left( \ell + 1 \right)} \Omega_3 \mathbf{I}_{2\times2} \, ,
\qquad \qquad \qquad \qquad
 \mathbf{\hat W}  =   \frac{15 }{\ell \left( \ell + 1 \right)}
  \left(\begin{array}{cc}
    0   &   \Omega_3 - 2 \omega_3   \\
    2 \varpi_3   &  0    \\
 \end{array}\right) \, .
\end{eqnarray}
here $\mathbf{I}_{2\times2}$ is the identity matrix of rank 2.  While
the matrices associated with the operator $\mathcal{M}$ are:
\begin{eqnarray}
 \mathbf{\hat M}_0   = \frac{1}{\ell \left( \ell + 1 \right)}
  \left(\begin{array}{cc}
    \ell \left( \ell + 1 \right)    & \quad   m r^2  e^{-2\lambda} \left( g_1 + 6 g_3 \right)   \\
    0    &  0
\\
 \end{array}\right) \,,  \qquad
 \mathbf{\hat M}_{1}  = \frac{1}{\ell \left( \ell + 1 \right)}
  \left(\begin{array}{cc}
     0  &  0 \\
     0  &  \quad  r^2 e^{-2\lambda} \left( g_1 + 6 g_3 \right)    \\
 \end{array}\right) \, ,
\end{eqnarray}
\begin{eqnarray}
 \mathbf{\hat M}_{2}  = \frac{m}{\ell \left( \ell + 1 \right)}
  \left(\begin{array}{cc}
    0      &  - \frac{15}{2}   r^2  e^{-2 \lambda} g_3     \\
    0      &   0    \\
 \end{array}\right) \,, \qquad  \qquad  \qquad \quad
 \mathbf{\hat M}_{3}   = \frac{1}{\ell \left( \ell + 1 \right)}
  \left(\begin{array}{cc}
     0  &  0 \\
     0  &  \quad - \frac{15}{2}  r^2 e^{-2\lambda} g_3     \\
 \end{array}\right)  \,  ,
\end{eqnarray}
where
\begin{eqnarray}
g_1  =  2 \left( \frac{1}{r} - \nu' \right) \varpi_1 + \varpi_1 '  \, ,  \qquad \qquad
g_3  =  2 \left( \frac{1}{r} - \nu' \right) \varpi_3 + \varpi_3 '  \, .
\end{eqnarray}

Finally, we consider the operators $\mathcal{Q}_{\sigma}^{\rm tot}$ and
$\mathcal{N}$ of Eq.~(\ref{eq:xi}), which exists only in the
non-barotropic case:
\begin{eqnarray}
\mathcal{Q}_{\sigma}^{\rm tot} & = &  \mathbf{\hat Q}_{\sigma} \mathcal{I}
                         ~\oplus~ \mathbf{\hat Q}_1 \mathcal{L}_{1}^{\pm 2} \, , \\
\mathcal{N}                & = &  \mathbf{\hat N}_0 \mathcal{I}  \, ,
\end{eqnarray}
where the matrices $\mathbf{\hat Q}_{\sigma}$, $\mathbf{\hat Q}_1$ and
$\mathbf{\hat N}_0$ have the form:
\begin{eqnarray}
 \mathbf{\hat Q}_{\sigma}  =
  \left(\begin{array}{cc}
 - \sigma + m  \left( \Omega_1 + 6 \Omega_3 \right)   & 0    \\
    0      &   0    \\
 \end{array}\right) \,, \qquad
 \mathbf{\hat Q}_{1}   =
  \left(\begin{array}{cc}
     - \frac{15}{2} \, m \, \Omega_3   &  0 \\
     0  &   0     \\
 \end{array}\right)  \,  ,
\qquad
 \mathbf{\hat N}_{0}   =
  \left(\begin{array}{cc}
     0   &  0 \\
     0  &   e^{2 \nu - 2 \lambda} \left( 1 - \frac{\Gamma_1}{\Gamma} \right)  \nu'     \\
 \end{array}\right)  \,  .   \label{eq:matQ}
\end{eqnarray}
However, the definition of
the vector $v^{\ell m}$ in~(\ref{def:usv}) and the values of the
matrix coefficients in~(\ref{eq:matQ}) lead to simpler equations:
\begin{equation}
\left[ - \sigma + m \left( \Omega_1 + 6 \Omega_3 \right) -
  \frac{15}{2} \, m \, \Omega_3 \mathcal{L}_{1}^{\pm 2} \right]
\xi^{\ell m} = e^{2 \nu - 2 \lambda} \left( 1 -
\frac{\Gamma_1}{\Gamma} \right) \nu' u_1^{\ell m} \, .
\end{equation}

\section{Angular Operators \label{Integrals}}

In this Appendix, we present the definition of the linear angular
operators $\mathcal{L}_{i}^{\pm j}$ with $i,j \in \mathbb{N}$, which
are used in the angular integration of the perturbation equations
derived in \cite{2007PhRvD..75f4019S} and  couple the various
perturbation functions. For convenience we don't  write down
the detailed expression of each operator, but rather the final part that
shows the couplings that each one of them introduces on a generic
harmonic component perturbation $\mathcal{A}_{\ell m}$.

The operators that introduce couplings of the form $\ell \pm 1$ are:
\begin{widetext}
\begin{eqnarray}
\mathcal{L}  _{1}^{\pm 1}  \mathcal{A}_{\ell m}& = & \left( \ell - 1 \right) Q_{\ell m} \mathcal{A}_{\ell-1 m}
                          - \left( \ell + 2 \right) Q_{\ell+1 m} \mathcal{A}_{\ell+1 m} \, , \\ \nn \\
\mathcal{L}  _{2}^{\pm 1}  \mathcal{A}_{\ell m}
& = & - \left( \ell+1 \right) Q_{\ell m} \mathcal{A}_{\ell-1 m} + l Q_{\ell+1 m} \mathcal{A} _{\ell+1 m} \, , \\ \nn \\
\mathcal{L}  _{3}^{\pm 1}  \mathcal{A}_{\ell m}
                   &=&  \left( \ell -1 \right) \left( \ell + 1 \right) Q_{\ell m} \mathcal{A}_{\ell-1 m}
                    +  \ell \left( \ell + 2 \right) Q_{\ell+1 m} \mathcal{A}_{\ell+1 m}   \, , \\ \nn \\
\mathcal{L}  _{4}^{\pm 1}  \mathcal{A}_{\ell m}
                   &=& Q_{\ell m} \mathcal{A}_{\ell-1 m}  + Q_{\ell+1 m} \mathcal{A}_{\ell+1 m}   \, , \\ \nn 
\end{eqnarray}
while the following operators introduce couplings of the form $\ell \pm 2$,
\begin{eqnarray}
\mathcal{L}  _{1}^{\pm 2}  \mathcal{A}_{\ell m}
 & = &   - Q_{\ell-1 m} Q_{\ell m} \mathcal{A}_{\ell-2 m}  + \left( 1 - Q_{\ell m}^{2} - Q_{\ell+1 m}^{2}\right)
                      \mathcal{A}_{\ell m} -  Q_{\ell m} Q_{\ell+1 m} \mathcal{A}_{\ell+2 m}    \label{eq:L1pm2}
   \, , \\ \nn \\
\mathcal{L}  _{2}^{\pm 2}  \mathcal{A}_{\ell m}
                   & = & - \left( \ell+1 \right) Q_{\ell-1 m} Q_{\ell m} \mathcal{A}_{\ell-2 m}
                     + \left[ \ell Q_{\ell+1 m}^{2} - \left( \ell+1 \right) Q_{\ell m}^{2} \right] \mathcal{A}_{\ell m}
                     + \ell Q_{\ell+1 m} Q_{\ell+2 m} \mathcal{A}_{\ell+2 m}   \, , \\ \nn \\
\mathcal{L}  _{3}^{\pm 2} \mathcal{A}_{\ell m}
                  & = &  \left( \ell-2 \right) Q_{\ell-1 m} Q_{\ell m} \mathcal{A}_{\ell-2 m}
                      + \left[ \ell Q_{\ell+1 m}^{2} - \left( \ell+1 \right) Q_{\ell m}^{2} \right] \mathcal{A}_{\ell m}
                      - \left( \ell +3 \right)  Q_{\ell+1 m} Q_{\ell+2 m} \mathcal{A}_{\ell+2 m}    \, ,  \\ \nn \\
\mathcal{L}  _{4}^{\pm 2}  \mathcal{A}_{\ell m}
                  & = & - \left(\ell-2\right) \left(\ell+1\right)
                  Q_{\ell-1 m}Q_{\ell m} \mathcal{A}_{\ell-2 m}
                  - \ell \left( \ell+3 \right) Q_{\ell+1 m} Q_{\ell+2 m}
                  \mathcal{A} _{\ell+2 m} \nn \\
                   && + \left[ \ell \left(\ell+1\right) - \left( \ell+1\right) \left( \ell-2\right) Q_{\ell m}^{2}
                      - \ell \left( \ell+3 \right) Q_{\ell+1 m}^{2} \right] \mathcal{A}_{\ell m} \, .
\end{eqnarray}
Finally, the following operators introduce couplings of the form $\ell \pm 3$,
\begin{eqnarray} 
\mathcal{L}  _{1}^{\pm 3}  \mathcal{A}_{\ell m}
                    & = &  - \left(\ell - 3 \right) Q_{\ell-2 m}  Q_{\ell-1 m} Q_{\ell m} \, \mathcal{A}_{\ell-3 m}
                       + \left(\ell +4 \right) Q_{\ell+1 m}  Q_{\ell+2 m} Q_{\ell+3 m} \, \mathcal{A}_{\ell+3 m}\nn \\
                     && + Q_{\ell m} \left[ \ell  Q_{\ell-1 m}^{2} + \left(\ell-1 \right) \left( 1 -  Q_{\ell m}^{2} -  Q_{\ell+1 m}^{2}
                        \right) \right]   \mathcal{A}_{\ell-1 m}
                       \nn  \\ &&
               - Q_{\ell+1 m}  \left[ \left( \ell + 1 \right)  Q_{\ell+2 m}^{2}  + \left( \ell+2\right)
                               \left(1 -  Q_{\ell m}^{2} - Q_{\ell+1 m}^{2} \right) \right]  \mathcal{A}_{\ell+1 m} \, ,
                      \\ \nn \\
\mathcal{L}  _{2}^{\pm 3}  \mathcal{A}_{\ell m}
            & = & - \left( \ell - 3\right) \left( \ell+1\right) Q_{\ell-2 m} Q_{\ell-1 m} Q_{\ell m} \mathcal{A}_{\ell-3 m}
              - \ell \left( \ell +4 \right) Q_{\ell+1 m} Q_{\ell+2 m} Q_{\ell+3 m} \mathcal{A}_{\ell+3 m} \nn \\
     &&  + Q_{\ell m} \left[  \ell \left( \ell+1 \right) Q^2_{\ell-1 m} - \left( \ell-1\right) \left( \ell+1\right) Q_{\ell m}^{2}
                     +  \ell \left( \ell-1 \right) Q^2_{\ell+1 m} \right] \mathcal{A}_{\ell-1 m} \nn \\
           &&   + Q_{\ell+1 m} \left[  \left( \ell+1 \right) \left( \ell+2 \right) Q^2_{\ell m}
                     - \ell \left( \ell+2\right) Q_{\ell+1 m}^{2} +  \ell \left( \ell+1 \right) Q^2_{\ell+2 m} \right]
                       \mathcal{A}_{\ell+1 m} \, , \\ \nn \\
\mathcal{L}  _{3}^{\pm 3}  \mathcal{A}_{\ell m}
                   & = &     \left( \ell+1\right) Q_{\ell-2 m} Q_{\ell-1 m} Q_{\ell m} \mathcal{A}_{\ell-3 m}
                    - \ell Q_{\ell+1 m} Q_{\ell+2 m} Q_{\ell+3 m} \mathcal{A}_{\ell+3 m} \nn \\
                  &&  - Q_{\ell m} \left[  \left( \ell+1 \right) + \ell Q^2_{\ell+1 m} - \left( \ell+1\right)
                        \left( Q_{\ell-1 m}^{2} + Q^2_{\ell  m} \right) \right] \mathcal{A}_{\ell-1 m} \nn \\
           &&  + Q_{\ell+1 m} \left[ \ell +  \left( \ell+1 \right) Q^2_{\ell m}
                       - \ell \left( Q_{\ell+1 m}^{2} + Q^2_{\ell+2 m}
                       \right) \right] \mathcal{A}_{\ell+1 m} \, ,
\end{eqnarray}
\end{widetext}
where $Q_{\ell m}$ is defined as
\begin{equation}
Q_{\ell m} \equiv \sqrt{\frac{\left( \ell-m \right) \left( \ell+m
\right)}{\left(2\ell-1\right) \left(2\ell+1\right)}} \, .
\label{eq:Qlm}
\end{equation}
A useful relation between operators is the following :
\begin{equation}
\mathcal{L} _{4}^{\pm 1} = -\frac{1}{2} ( \mathcal{L} _{1}^{\pm 1} +
\mathcal{L} _{2}^{\pm 1} ) \nonumber \, .
\end{equation}

\nocite*

\end{document}